\documentclass[preprint,12pt]{elsarticle}

\usepackage{amssymb}
\usepackage{bm}

\journal{Nuclear Physics A}

\begin{document}

\begin{frontmatter}



\title{Double-pole nature of $\Lambda(1405)$ studied with coupled-channel 
complex scaling method using complex-range Gaussian basis}


\author[KEK,JPARCb]{A. Dot\'e}
\ead{dote@post.kek.jp}

\author[OIT,RCNP]{T. Myo}

\address[KEK]{KEK Theory Center,
Institute of Particle and Nuclear Studies (IPNS), 
High Energy Accelerator Research Organization (KEK), 1-1 Oho, Tsukuba, Ibaraki, 305-0801, Japan}

\address[JPARCb]{J-PARC Branch, KEK Theory Center, IPNS, KEK,
203-1, Shirakata, Tokai, Ibaraki, 319-1106, Japan}

\address[OIT]{General Education, Faculty of Engineering, Osaka Institute of Technology, Osaka 535-8585, Japan}

\address[RCNP]{Research Center for Nuclear Physics (RCNP), Osaka University, Ibaraki 567-0047, Japan}

\begin{abstract}

The excited hyperon $\Lambda(1405)$ is the important building 
block for kaonic nuclei which are nuclear many-body system 
with anti-kaons. We have been investigating the $\Lambda(1405)$ 
resonance with the coupled-channel Complex Scaling Method (ccCSM) 
in which the $\Lambda(1405)$ is treated as a hadron-molecular state of a 
$\bar{K}N$-$\pi\Sigma$ coupled system. We use a $\bar{K}N$(-$\pi Y$) potential 
based on the chiral SU(3) theory. 
In this article, we report the double-pole nature of the 
$\Lambda(1405)$, which is a characteristic property predicted by 
many studies with chiral SU(3)-based models. 
With the help of the complex-range Gaussian basis in ccCSM, 
we have found successfully another pole besides 
a pole near the $\bar{K}N$ threshold  (called higher pole) 
which was found in our previous work with the real-range Gaussian basis. 
The new pole (called lower pole) is found far below $\bar{K}N$ threshold: 
$(M, -\Gamma/2)=(1395, -138)$ MeV when $f_\pi=110$ MeV. 
In spite of so broad width of the lower-pole state, 
the state is clearly identified with good separation from continuum states, 
since the oscillatory behavior of the continuum states is improved 
owing to the complex-range Gaussian basis. 
Analyzing the ccCSM wave function of the lower pole, 
we have revealed explicitly that the lower-pole state is dominated by 
the $\pi\Sigma$ component rather than the $\bar{K}N$ component. 
We have confirmed that the ccCSM wave function is correctly connected 
to the asymptotic form of the resonance wave function. 
Estimating the meson-baryon mean distance for the lower-pole state 
which involves a large decay width, the obtained value has 
a large imaginary part comparable to a real part. 
Therefore, the mean-distance of the lower-pole state is difficult 
to be interpreted intuitively. Such a nature of the lower pole is 
different from that of the higher pole. 
In addition, we have investigated the origin of the appearance 
of the lower pole. The lower pole is confirmed to be generated 
by the energy dependence attributed to the chiral dynamics, 
by comparing the result of an energy-independent potential.
\end{abstract}


\end{frontmatter}


\section{Introduction}
\label{}

Nuclear system with anti-kaons ($\bar{K}$ = $K^-$, $\bar{K}^0$ mesons) 
has been an important topic in strange nuclear physics and hadron physics, 
since the strong attraction between anti-kaon and nucleon is expected to 
provide a dense nuclear state \cite{AY_2002, AMDK}. 
To know the detailed property of such an interesting system, 
we have investigated finite nuclear system with anti-kaon, which is called 
as kaonic nuclei. Many works have been devoted to the 
study of a three-body system of ``$K^-pp$'' (two nucleon with 
an anti-kaon), which is a prototype of kaonic nuclei 
\cite{Faddeev:Ikeda, Faddeev:Shevchenko, Kpp:DHW, Kpp:Wycech, Kpp:Barnea}. 
According to the past studies, the binding energy and decay width of 
$K^-pp$ depend on theoretical approaches; the $K^-pp$ binding energy is 
ranging from 20 to 90 MeV, and decay width is more than 40 MeV 
\cite{SummaryKpp}. 
Therefore, the conclusive result has not been achieved in theoretical 
studies. However, all theoretical studies commonly suggest that 
$K^-pp$ exists as a resonant state between $\bar{K}NN$ and $\pi\Sigma N$ thresholds and that the coupling of $\bar{K}N$ and $\pi\Sigma$ is 
important for this system.  

Motivated by such current situation of theoretical studies, 
we have started a project to investigate kaonic nuclei 
with {\it coupled-channel Complex Scaling Method ``ccCSM''}. 
The ccCSM is a suitable method to investigate 
the kaonic nuclei since this method can treat simultaneously 
two important ingredients in the study of the kaonic nuclei: 
1. coupled-channel problem and 2. resonant state. 
In particular, it has been well confirmed in many studies of 
unstable nuclei that the complex scaling method is quite effective 
to investigate resonant states \cite{CSM:Myo}. 
As the first step of our attempt, 
we applied the ccCSM to a two-body $\bar{K}N$-$\pi Y$ coupled 
system \cite{ccCSM-KN_NPA}. Dealing with both resonance 
and scattering problems in the single framework of the ccCSM,  
we studied the scattering amplitude of the $\bar{K}N$-$\pi Y$ 
system and the property of the resonance in the $I=0$ channel 
which corresponds to the $\Lambda(1405)$. 

In this article, we make a deeper consideration on 
the $I=0$ resonance pole of $\bar{K}N$-$\pi\Sigma$ system 
because of following reasons.  
Resonance poles are important components of scattering states, 
although the pole itself is not a direct observable measured by experiments.  
If resonance poles are well understood in a scattering system, 
the scattering state can be decomposed into resonant and 
non-resonant components. We can extract resonance nature from 
the scattering system by performing such analysis. 
In particular, using the complex scaling method we can discuss 
on the resonance pole in the same way as bound states, 
since the CSM provides a resonance wave function explicitly. 
In past studies of the $\bar{K}N$-$\pi\Sigma$ system, 
the resonance nature was investigated via the analysis of 
the wave function on the {\it real-energy} axis \cite{ChU:KSW, 
L1405wfn:Takeuchi, L1405wfn:Yamagata}. 
On the other hand, we investigate directly the wave function of 
resonance-pole state. In other words, we analyze the resonance 
wave function at the pole on the {\it complex-energy} plane. 

In the $I=0$ $\bar{K}N$-$\pi\Sigma$ system, 
there is an interesting nature which is so-called {\it double-pole structure} 
of $\Lambda(1405)$. 
We use a $\bar{K}N$-$\pi\Sigma$ potential which is based on the chiral 
SU(3) theory \cite{ChU:KSW}, 
since the pion and anti-kaon of Nambu-Goldstone bosons 
follow the chiral dynamics. 
It was pointed out in Refs. \cite{ChU:OM, ChU:Jido} that 
there appear two poles in the $I=0$ channel of the $\bar{K}N$-$\pi\Sigma$ 
system when the $\Lambda(1405)$ resonance is considered 
with a chiral SU(3)-based interaction. 
One pole (called higher pole) appears close to the $\bar{K}N$ threshold 
involving a small decay width, while the other pole (called lower pole) 
appears far below the $\bar{K}N$ threshold involving a large decay width. 
In the further study \cite{ChU:HW}, such a double-pole structure 
of $\Lambda(1405)$ was confirmed in several models of 
coupled-channel chiral dynamics \cite{ChU:ORB, ChU:HNJH, ChU:BNW, ChU:BMN}. 
They showed that the higher pole is originated from a $\bar{K}N$ bound state, 
whereas the lower pole comes from a $\pi\Sigma$ resonance state. 
As a result of these studies, the higher-pole position is well established 
to be around $(M, -\Gamma/2)=(1420, -20)$ MeV, although the lower-pole 
position depends strongly on models.  
The decay width of the lower pole is obtained to 
be quite large compared to the higher pole. 
It is noted that some of latest studies \cite{ChMM:CS, ChU:MM, ChU:GO} 
report slightly different positions of the double pole 
from those of earlier studies. 
Anyway, such a double-pole structure is generated by the energy dependence 
of the chiral SU(3)-based interaction \cite{KN-pS:IHJ}. 
Therefore, it is considered that 
the double-pole structure of $\Lambda(1405)$ 
is an important consequence of the chiral dynamics. 

In our previous study with ccCSM \cite{ccCSM-KN_NPA}, since the lower pole 
was not so clearly found, we were not able to conclude its existence. 
The pole that we found in the $I=0$ channel 
corresponds to the higher pole, since its complex energy is similar to 
the known value of the higher-pole energy. 
The lower pole is expected to exist, because our potential is 
based on the chiral theory.  
In our previous analysis, it was hard to find a resonance pole 
with a broad width such as the lower pole 
by means of the complex scaling method with an ordinary real-range 
Gaussian basis that we used, since a broad resonance is not 
easily separated from the continuum states, because of 
the limitation of the real-range Gaussian basis 
for describing the oscillatory behavior of the wave function. 
Recently, Kamimura {\it et al.} proposed the use of the complex-range 
Gaussian basis in the complex scaling method, instead of the real-range 
Gaussian basis which has been used so far. They applied this new basis to 
the description of resonant states of $^{12}$C with $3\alpha$-cluster model. 
They succeeded in finding broad resonances 
which were difficult to be found with the usual complex scaling method 
using the real-range Gaussian basis.
\cite{CSM-CG}. 

In this article, we apply the complex-range Gaussian basis to our 
coupled-channel Complex Scaling Method, in order to find clearly 
the lower pole of the $I=0$ $\bar{K}N$-$\pi Y$ system 
and investigate its wave function. 
The contents of this article is as follows. 
We explain our formalism, in particular, on the complex-range 
Gaussian basis in the section 2. In the section 3, we show the result 
of our calculation and make an analysis of our result. 
We summarize the present study and mention 
on our future plan with this method in the section 4.

\section{Formalism}
\label{Formalism}

We are investigating resonance states of the $\bar{K}N$-$\pi\Sigma$ coupled system with the $s$-wave and isospin zero by means of the coupled-channel Complex Scaling Method (ccCSM). In the ccCSM, Hamiltonian $\hat{H}$ is complex-scaled: $\bm{r} \rightarrow \bm{r}e^{i\theta}$ and $\bm{k} \rightarrow \bm{k}e^{-i\theta}$, 
$\bm{r}$ and $\bm{k}$ are the coordinate and its conjugate momentum, 
respectively. 
When the complex-scaled Hamiltonian $\hat{H}_\theta$ is diagonalized with a 
square-integrable function, complex eigenvalues 
for continuum states distribute along with lines $\tan^{-1} ({\rm Im} \, E \, / \, {\rm Re} \, E) = -2\theta$ running from thresholds 
(so-called $2\theta$ line) on the complex-energy plane. 
On the other hand, complex eigenvalues for resonance states appear 
off the $2\theta$ line and they are independent of the scaling angle $\theta$. 
Then, we can find the resonant states. Detailed explanation of the ccCSM is given in Ref. \cite{ccCSM-KN_NPA}.

In the studies of resonances with the complex scaling method, the Gaussian 
basis function 
has been often employed because of its convenience for 
the analytical calculation of matrix elements. 
In the present study, we use the Gaussian functions with 
$complex$ range parameters as proposed by Kamimura {\it et al.} \cite{CSM-CG}:
\begin{eqnarray}
G^{(\pm)}_n (r) \; = \; N^{(\pm)}_n \; r^l \, \exp[-(a_n \pm i\omega ) \, r^2],
\end{eqnarray} 
where $N^{(\pm)}_n$ means a normalization factor, $\{a_n\}$ is a real 
range parameter, and  $\pm i\omega$ is introduced to make the range parameters complex. 
Considering the linear combination of the basis functions 
$\{G^{(\pm)}_n (r)\}$, they are 
equivalent to Gaussian functions with trigonometric functions: 
\begin{eqnarray}
G^{(\cos)}_n (r) &=& N^{(\cos)}_n \; r^l \, \exp[-a_n \, r^2] \cos \omega r^2 
\; \sim \; G^{(+)}_n (r) +  G^{(-)}_n (r), \label{Eq:Gauss-cos_CSM}\\
G^{(\sin)}_n (r) &=& N^{(\sin)}_n \; r^l \, \exp[-a_n \, r^2] \sin \omega r^2 
\; \sim \; G^{(+)}_n (r) -  G^{(-)}_n (r). \label{Eq:Gauss-sin_CSM}
\end{eqnarray} 

Compared with the real-range Gaussian basis which has been used so far 
in the complex scaling method, 
the complex-range Gaussian basis describes better the spatially oscillatory 
behavior of eigenstates for the complex-scaled Hamiltonian 
as expected by its trigonometric nature. 

To find resonant states with large decay widths, the description of 
continuum states is also important, because such broad resonances tend to 
mix with the continuum states. In case of the real-range Gaussian basis, 
the eigenvalues of $\hat{H}_\theta$ for the continuum states often 
deviate from $2\theta$ line gradually with increase of the imaginary energy. 
This is due to the limitation for description of the spatial oscillation. 
In addition, we need complex-rotation of the Hamiltonian with large angle 
$\theta$, when we search for such broad resonances. However, 
the oscillation of the resonance wave function is more enhanced, getting 
larger $\theta$. Therefore, it becomes difficult to find 
broad resonances with the real-range Gaussian basis. 
With help of an analytic continuation in the coupling constant, 
people has evaluated the complex energy of broad resonances 
in the complex scaling method. (ACCC+CSM \cite{ACCC+CSM:Aoyama}) 
Using the complex-range Gaussian basis, we expect that resonances 
with broad decay width can be clearly obtained. 

Other set up for the present study of the $s$-wave and 
isoscalar $\bar{K}N$-$\pi\Sigma$ system is the same as 
that for our previous study \cite{ccCSM-KN_NPA}. 
The wave function of $s$-wave and isoscalar $\bar{K}N$-$\pi\Sigma$ system is 
given as 
\begin{eqnarray}
| \Phi^{I=0, l=0}_{\bar{K}N-\pi\Sigma} \rangle 
& = & \frac{1}{r}\phi^{l=0}_{\bar{K}N}(r) \, Y_{00}(\Omega) | \bar{K}N \rangle  \; + \;  
\frac{1}{r} \phi^{l=0}_{\pi\Sigma}(r) \, Y_{00}(\Omega) | \pi\Sigma \rangle. 
\label{Eq:I=0wfnc}
\end{eqnarray}
The spatial part of wave function, $\phi^{l=0}_{\alpha}(r)$, 
is complex-scaled in the ccCSM. 
We expand the complex-scaled wave function $\phi^{l=0, \, \theta}_{\alpha}(r)$ with the complex-range Gaussian basis,
\begin{eqnarray}
\phi^{l=0, \, \theta}_{\alpha}(r) & = & 
\sum_{n=1}^N \left\{ C^{\alpha, \, \theta}_{(\cos), n} G^{(\cos)}_n(r) \, + \, 
C^{\alpha, \, \theta}_{(\sin), n} G^{(\sin)}_n(r) \right\}. \label{Eq:GaussE_CSM}
\end{eqnarray} 
The complex parameters $\{C^{\alpha, \, \theta}_{(\cos), n}, \, C^{\alpha, \, \theta}_{(\sin), n}\}$ are determined by diagonalization of the complex-scaled Hamiltonian $\hat{H}_\theta$. 
The real-range parameter for these basis functions, which is described as $\{a_n\}$ 
in Eqs. (\ref{Eq:Gauss-cos_CSM}) and (\ref{Eq:Gauss-sin_CSM}), 
is generated by geometric progression:  
$a_n=1/b_n^2$ and $b_n=b_1 \gamma^{n-1}$. In this study, the parameters $b_1$ and $\gamma$ 
are taken to be 0.2 fm and 1.2, respectively. 
As for the imaginary-range parameter $\omega$, we will mention in 
the next section. 

We remark on the normalization and the norm of resonances in the ccCSM. 
The total wave function given in Eq. (\ref{Eq:I=0wfnc}) is normalized to 
unity when it is complex-scaled. As a result, norms of all channels 
are summed up to be 1; 
$\sum_\alpha \langle \tilde{\phi}^{l=0, \theta}_\alpha | 
\phi^{l=0, \theta}_\alpha \rangle =1.$ 
On the other hand, the norm of each component can be {\it a complex value} and 
{\it exceed the unity} in our study as shown in latter sections, since 
the ccCSM is based on Non-Hermitian quantum mechanics \cite{NH-QM:Moiseyev}, 
differently from the standard quantum mechanics that treats Hermitian. 
Although the probabilistic interpretation is no longer available here, 
we refer the norm in our analysis since 
it is a useful indicator to know major/minor components. 

As for Hamiltonian, we use a chiral SU(3)-based potential with Gaussian form given in the coordinate space which we proposed in our previous paper 
\cite{ccCSM-KN_NPA}. 
Since we consider the non-relativistic kinematics in this paper, 
we employ the non-relativistic versions of our potential (``NRv1'' and ``NRv2'').  
Here, we remark on our potential. Since it is an energy-dependent 
potential, we need to consider the self-consistency for the complex energy 
when we search for resonance poles. Such self-consistency is correctly 
treated in the present study as explained in detail in the previous paper.

\section{Result}
\label{}

\subsection{Lower pole of the $I=0$ $\bar{K}N$-$\pi\Sigma$ system}
\label{}

\begin{figure}
  \centering
  \includegraphics[width=.55\textwidth]{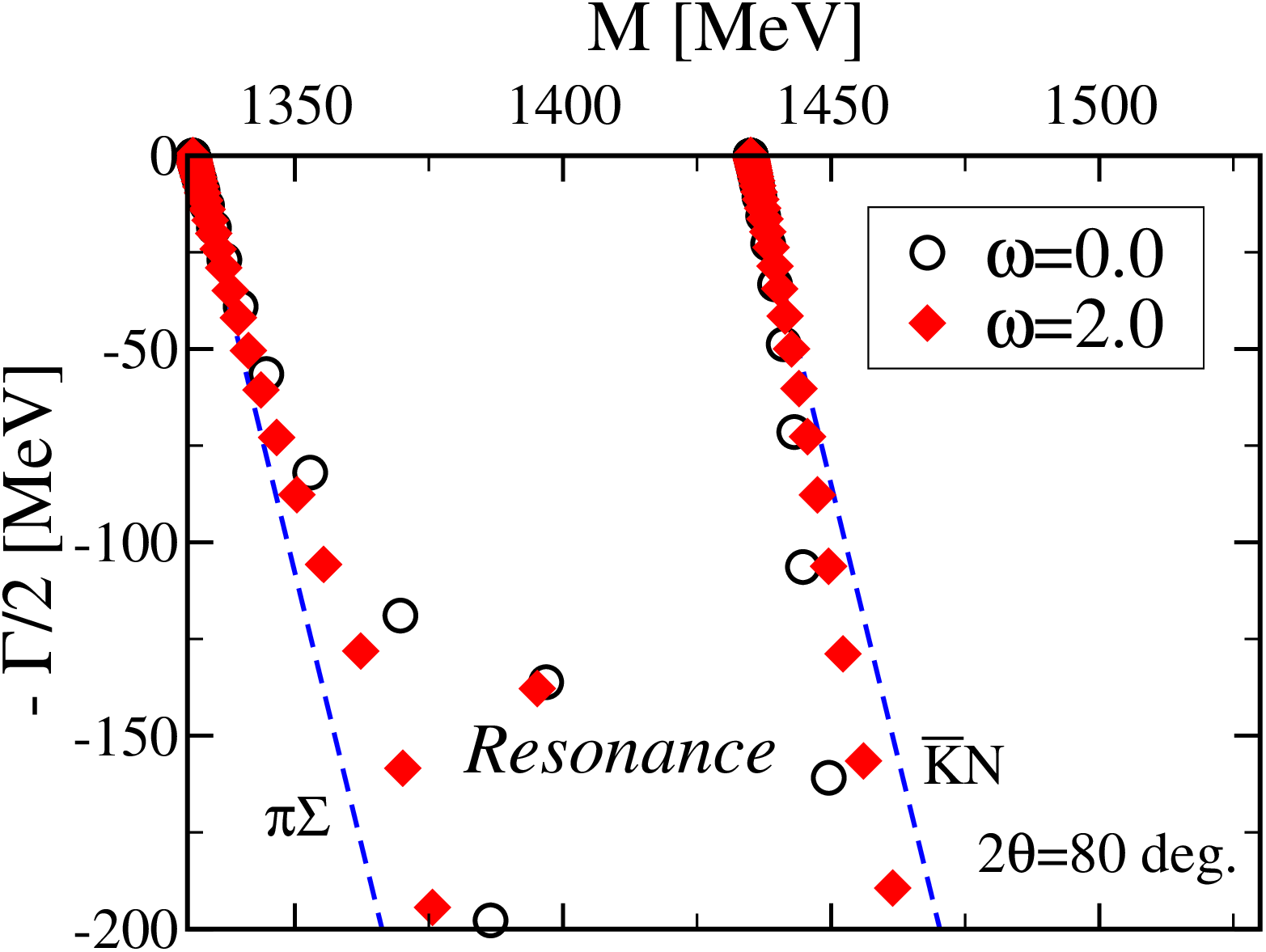}
\caption{Distribution of complex eigenvalues calculated with 
the NRv2 potential ($f_\pi=110$ MeV), involving 
the lower pole (``$Resonance$'').  
The scaling angle $\theta$ is set to 
be $40^\circ$. Red diamonds and black circles are eigenvalues calculated with 
complex-range Gaussian basis ($\omega=2.0$) and real-range one ($\omega=0.0$), respectively. Dashed lines are $2\theta$ lines indicating the $\bar{K}N$ and 
$\pi\Sigma$ continuum states. 
\label{EGVs_fp=110}}
\end{figure}

We can find the lower pole of $\bar{K}N$-$\pi\Sigma$ system in $I=0$ channel 
with the coupled-channel Complex Scaling Method. Using the complex-range 
Gaussian basis, we have obtained clearly another self-consistent solution of 
the energy-dependent Hamiltonian as well as the solution corresponding 
to the higher pole which has been reported in the previous work
\cite{ccCSM-KN_NPA}.

Fig. \ref{EGVs_fp=110} shows the distribution of the complex-energy eigenvalues  obtained for $f_\pi=110$ MeV case. To identify the lower pole clearly, we need the complex scaling with a large angle as $40^\circ$, since the lower pole has large imaginary energy compared with the real energy. Certainly, with a real-range Gaussian basis, we obtain a self-consistent solution which corresponds to a lower pole. However, since eigenvalues belonging to the continuum states are deviated from the $2\theta$ lines as depicted with black circles in Fig. \ref{EGVs_fp=110}, we can not conclude the existence of the solution of the lower pole. 
When we use the complex-range Gaussian basis, description of continuum states is improved. Their eigenvalues appear well along with the $2\theta$ lines. (Red diamonds in Fig. \ref{EGVs_fp=110}) 
The lower pole is confirmed to be prominently separated from the continuum states. 

\begin{table}
\caption{$\omega$ dependence of lower and higher poles at $\theta=40^\circ$. 
All poles are obtained to satisfy the self-consistency for the complex energy. 
$(M, -\Gamma/2)$ means the complex energy of the pole. 
The potential NRv2 ($f_\pi=110$ MeV) is used. 
\label{OmegaTra-NRv2}}
\begin{tabular*}{\textwidth}{l@{\hspace{0.5cm}}|@{\hspace{0.7cm}}r@{\hspace{0.7cm}}r@{\hspace{0.9cm}}r@{\hspace{0.9cm}}r@{\hspace{0.9cm}}r}
\hline
$\omega$ & 0.0 & 0.5 & 1.0 & 1.5 & 2.0 \\
\hline
Lower pole&&&&& \\
$M$         & 1396.9   & 1395.2   & 1395.1   & 1395.1  & 1395.2   \\
$-\Gamma/2$ & $-136.9$ & $-138.4$ & $-138.0$ & $-137.9$ & $-137.9$ \\
\hline
Higher pole&&& \\
$M$         & 1417.6   & 1417.9  & 1417.9  & 1417.9  & 1417.9  \\
$-\Gamma/2$ & $-16.3$  & $-16.6$ & $-16.6$ & $-16.6$ & $-16.6$ \\
\hline
\end{tabular*}
\end{table}

\begin{table}
\caption{$\theta$ trajectory of lower pole with $\omega=2.0$. 
The scaling angle $\theta$ is given in unit of degree. 
The potential NRv2 ($f_\pi=110$ MeV) is used. 
\label{ThetaTra-NRv2}}
\begin{tabular*}{\textwidth}{l@{\hspace{0.7cm}}|@{\hspace{0.5cm}}r@{\hspace{0.7cm}}r@{\hspace{0.7cm}}r@{\hspace{0.7cm}}r@{\hspace{0.7cm}}r@{\hspace{0.7cm}}r}
\hline
$\theta$ & 30  & 35  & 37  & 39  & 40  & 42  \\
\hline
$M$         & 1415.0 & 1398.7 & 1395.6 & 1395.2 & 1395.2 & 1395.2 \\
$-\Gamma/2$ &  $-122.3$ &  $-133.4$ &  $-136.6$ &  $-137.7$ &  $-137.9$ &  $-137.8$\\
\hline
\end{tabular*}
\end{table}

\begin{table}
\caption{Complex energy of lower and higher poles for each $f_\pi$ value. 
Both poles are calculated with the complex-range Gaussian basis ($\omega=2.0$) 
at $\theta=40^\circ$. Values of higher poles in parentheses are 
obtained with the real-range Gaussian basis at $\theta=30^\circ$. 
The potential NRv2 is used. 
\label{PolePositions-NRv2}}
\begin{tabular*}{\textwidth}{ll@{\hspace{0.5cm}}|@{\hspace{0.5cm}}r@{\hspace{1.0cm}}r@{\hspace{1.0cm}}r@{\hspace{1.0cm}}r}
\hline
 &$f_\pi$ & 90 & 100 & 110 & 120 \\
\hline
Lower pole&&&&& \\
 &$M$ & $\sim$1350 & 1368.8 & 1395.2 & 1424.7 \\
 &$-\Gamma/2$ & $\sim$ $-66$ & $-109.1$ & $-137.9$ &  $-163.2$ \\
\hline
Higher pole&&&&& \\
 &$M$ &  1419.9  &  1418.0  &  1417.9  &  1418.4 \\
 &    & (1419.9) & (1418.0) & (1417.8) & (1418.3) \\
 &$-\Gamma/2$ &  $-23.1$  &  $-19.8$  &  $-16.6$  &   $-14.1$ \\
 &            & ($-23.1$) & ($-19.8$) & ($-16.6$) &  ($-14.0$) \\
\hline
\end{tabular*}
\end{table}

\begin{figure}
  \centering
  \includegraphics[width=.55\textwidth]{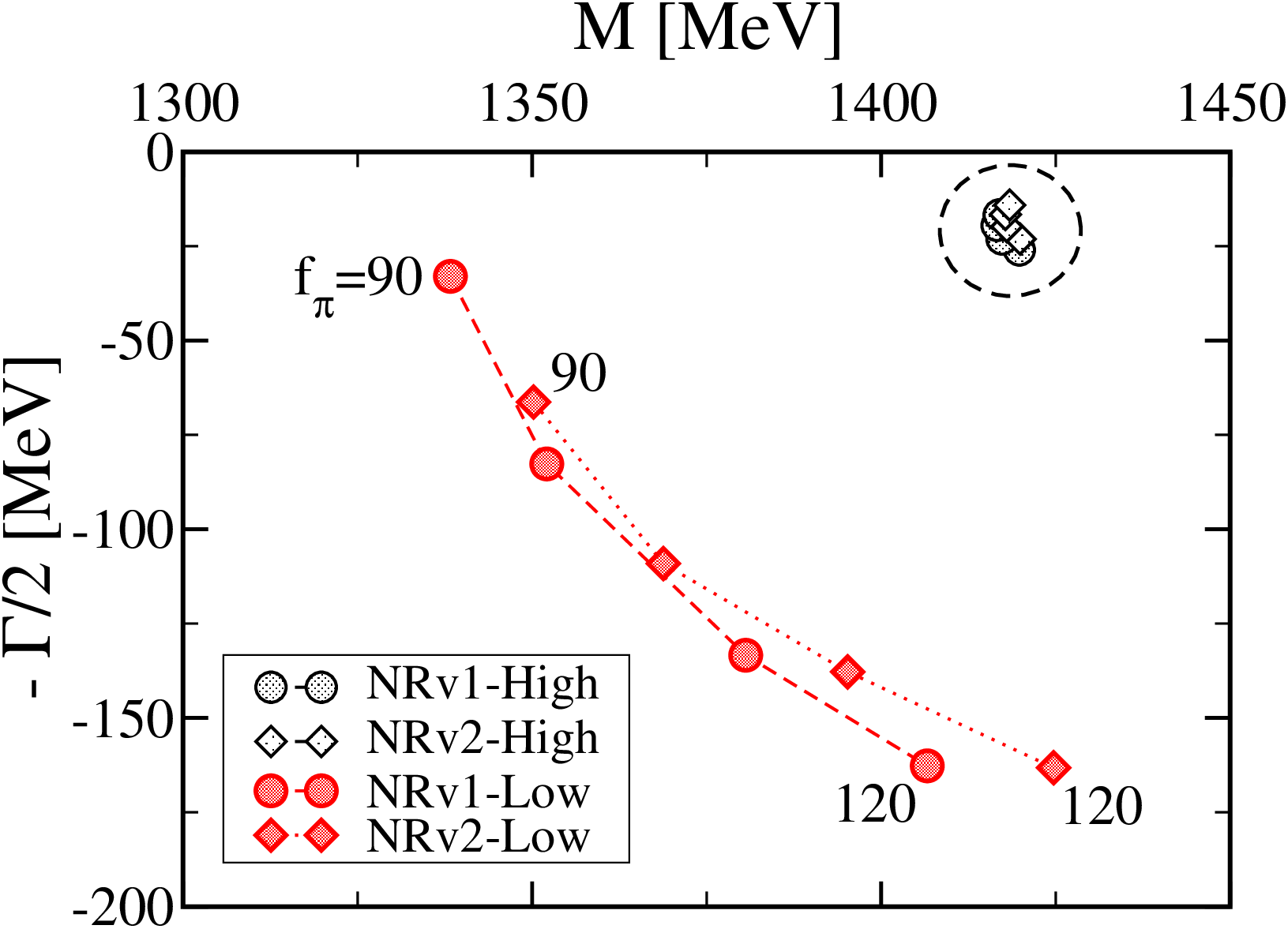}
\caption{Lower and higher poles obtained with NRv1 and NRv2 potentials. 
($M$, $-\Gamma/2$) means the complex energy of the pole in unit of MeV. 
$f_\pi$ value is varied from 90 to 120 MeV along with dashed line. 
Higher poles are concentrated in the area enclosed by 
a dashed-line circle. 
Both poles are calculated at the scaling angle $\theta=40^\circ$. 
\label{DP-NRv2-mfpi}}
\end{figure}

We have checked the stability of the lower pole. 
As shown in Table \ref{OmegaTra-NRv2}, it is confirmed that 
the complex eigen energy of the lower pole is almost independent of 
the imaginary part of the complex range, $\omega$, of the Gaussian basis. 
The $\theta$ trajectory has been also checked as shown in 
Table \ref{ThetaTra-NRv2}. Since the lower pole has 
large imaginary energy, its complex energy becomes stable at large 
angle around $\theta=40^\circ$. 
It is also confirmed that 
the lower-pole energy is stable for the number of basis $N$ in 
Eq. (\ref{Eq:GaussE_CSM}). 
$N$ is taken to be 40 in usual calculation, but it is increased upto 130 
to check the numerical stability of the lower pole. 
In addition, we have confirmed that the higher pole is very stable 
in all trajectories because of the small imaginary energy of the state. 

In Table \ref{PolePositions-NRv2}, we summarize the complex energy 
of lower and higher poles calculated with the NRv2 potential when 
$f_\pi$ value is varied from 90 to 120 MeV. The higher-pole energy is 
not so dependent on $f_\pi$ value, while the lower-pole one 
is found to change drastically depending on $f_\pi$ value. 
As explained in the previous work \cite{ccCSM-KN_NPA}, the potential used here 
is constrained by the 
$\bar{K}N$ scattering length at each $f_\pi$ value. 
Since the higher pole appears close to 
the $\bar{K}N$ threshold, its property is governed 
by the $\bar{K}N$ dynamics. Actually, the dominant component of the 
higher pole is found to be the $\bar{K}N$ component as is shown in 
the next section.  
On the other hand, the lower pole appears far from the $\bar{K}N$ threshold, 
rather close to the $\pi\Sigma$ threshold. Hence the lower pole 
is considered to be related to the $\pi\Sigma$ dynamics. 
Since the present potential is determined under the constraint for $\bar{K}N$ 
observable, not $\pi\Sigma$ one, the potential does not restrict so strongly 
the lower-pole property. 
Therefore, as shown in Fig. \ref{DP-NRv2-mfpi}, 
on the complex-energy plane the higher pole appears 
in the narrow area (enclosed by the dashed-line circle in the figure) 
independently of $f_\pi$, whereas 
the lower pole moves in the wide region, depending on the $f_\pi$ value. 

We note that the higher-pole energy is very stable with respect to 
the choice of the basis functions. 
In Table \ref{PolePositions-NRv2}, the values with parentheses 
on the higher-pole column are those obtained with the real-range Gaussian 
basis at $\theta=30^\circ$ \cite{ccCSM-KN_NPA}. 
The higher-pole energy is confirmed to be quite the same 
between both present and previous calculations. 
Therefore, in this article the higher pole is 
calculated in the same condition as for the lower pole, 
namely by the complex-range Gaussian basis with 
$\omega=2.0$ at $\theta=40^\circ$.

\subsection{Properties of the lower pole}
\label{Prop_Lower}

We investigate the property of the lower-pole state which is obtained with the ccCSM using the complex-range Gaussian basis. In this section, the case of the NRv2 potential with $f_\pi=110$ MeV is discussed as a typical result of our calculation. 

\begin{figure}
\centering
  \includegraphics[width=.41\textwidth]{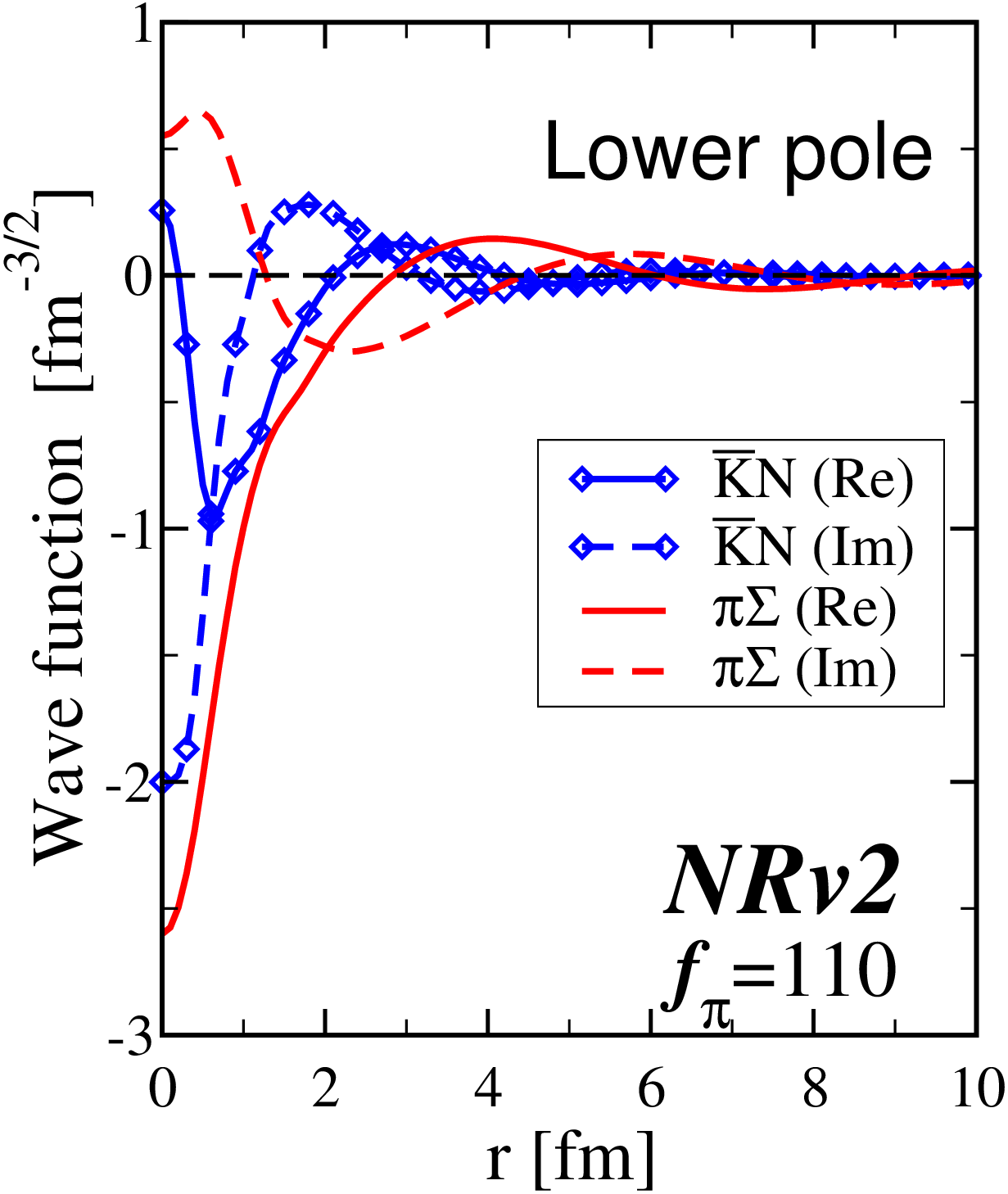}
  \hspace{0.8cm}
  \includegraphics[width=.42\textwidth]{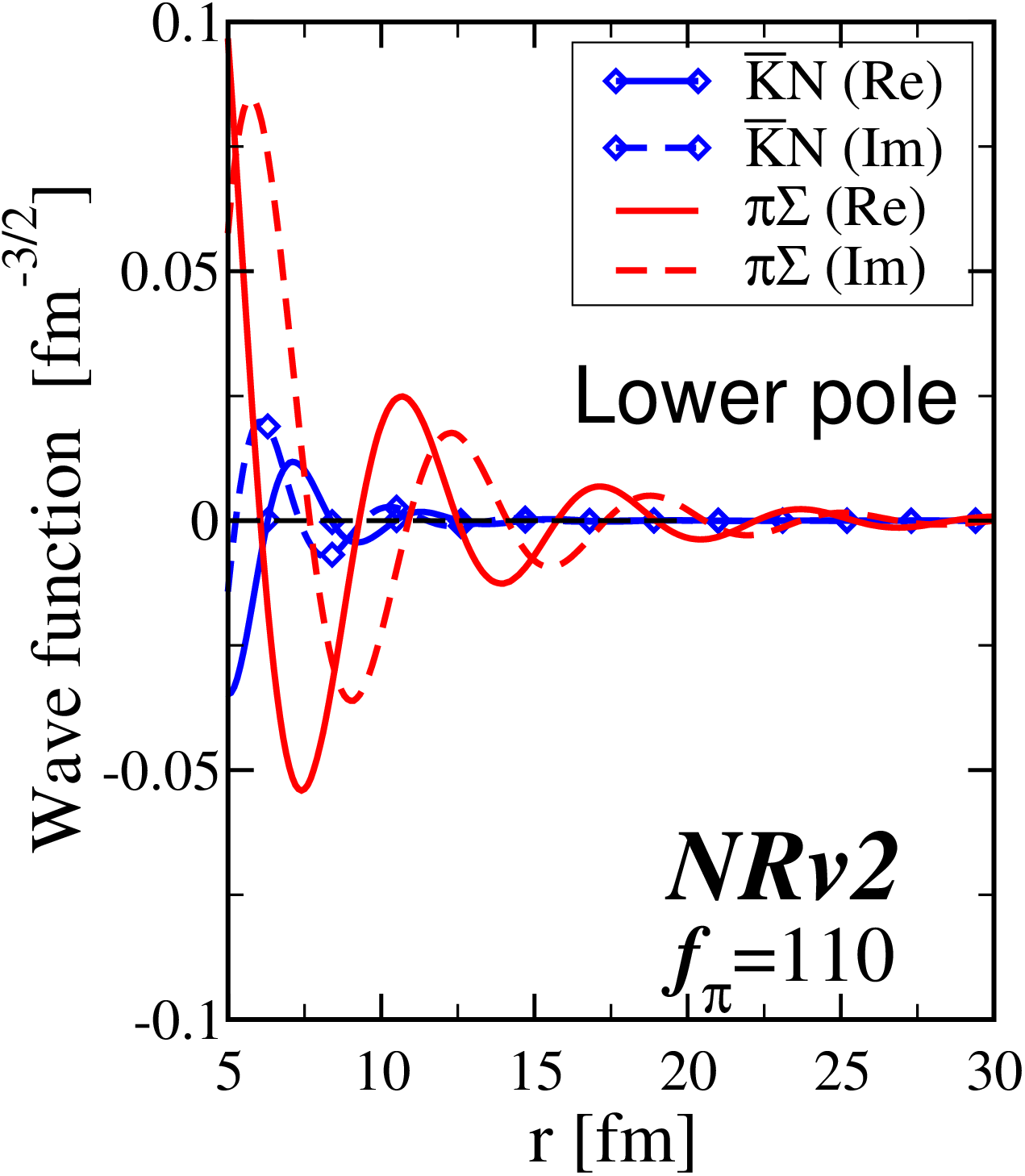}
  \caption{The ccCSM wave function of the lower-pole state 
of the NRv2 potential ($f_\pi=110$ MeV). 
Blue (red) lines show $\bar{K}N$ ($\pi\Sigma$) wave function, whose 
real (imaginary) part is drawn with solid (dashed) lines. 
Left: $r=0\sim10$ fm, right: $r=5\sim30$ fm. 
\label{NRv2-Lower_wfnc}}
\end{figure}

\begin{figure}
\centering
  \includegraphics[width=.41\textwidth]{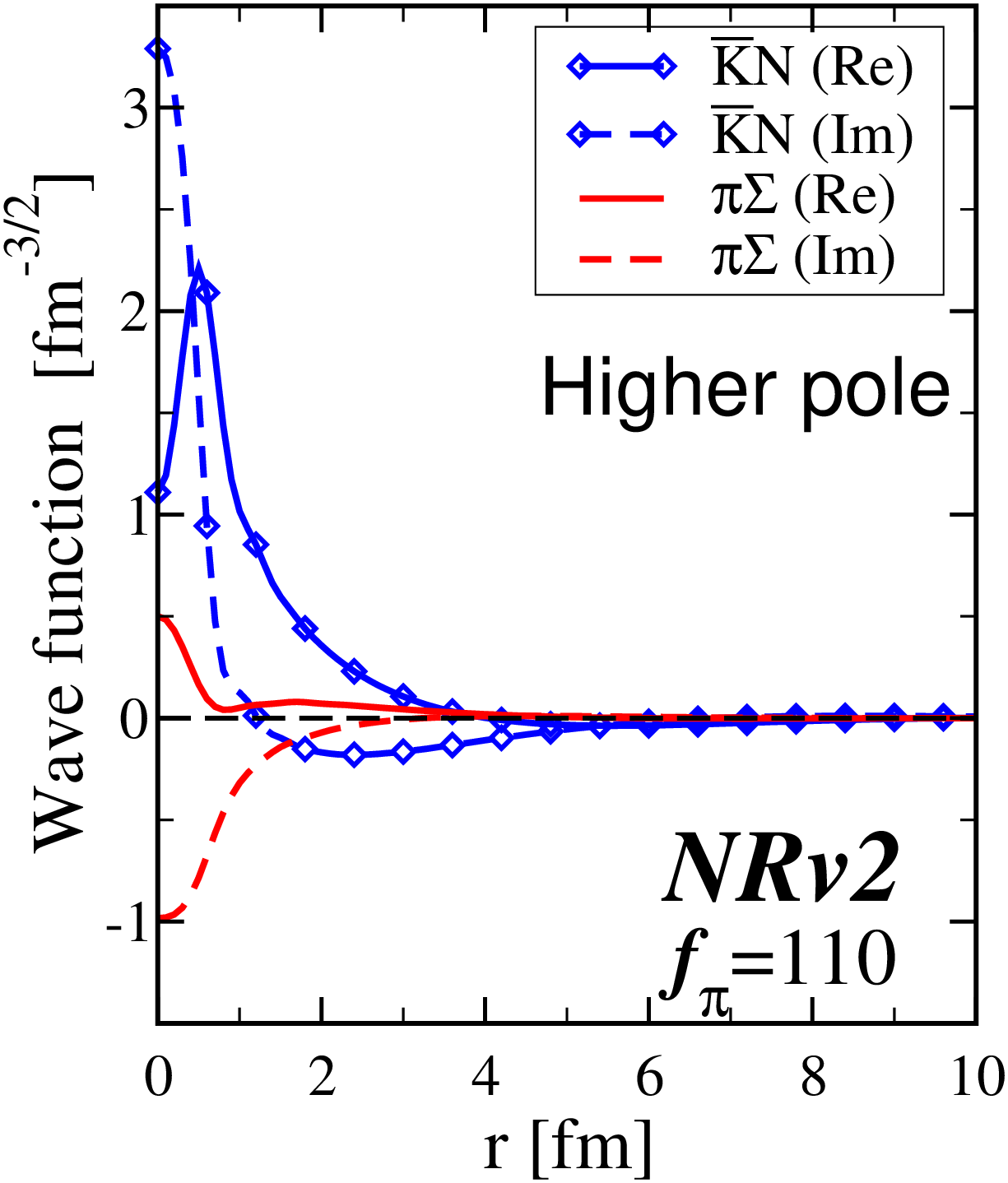}
  \hspace{0.8cm}
  \includegraphics[width=.42\textwidth]{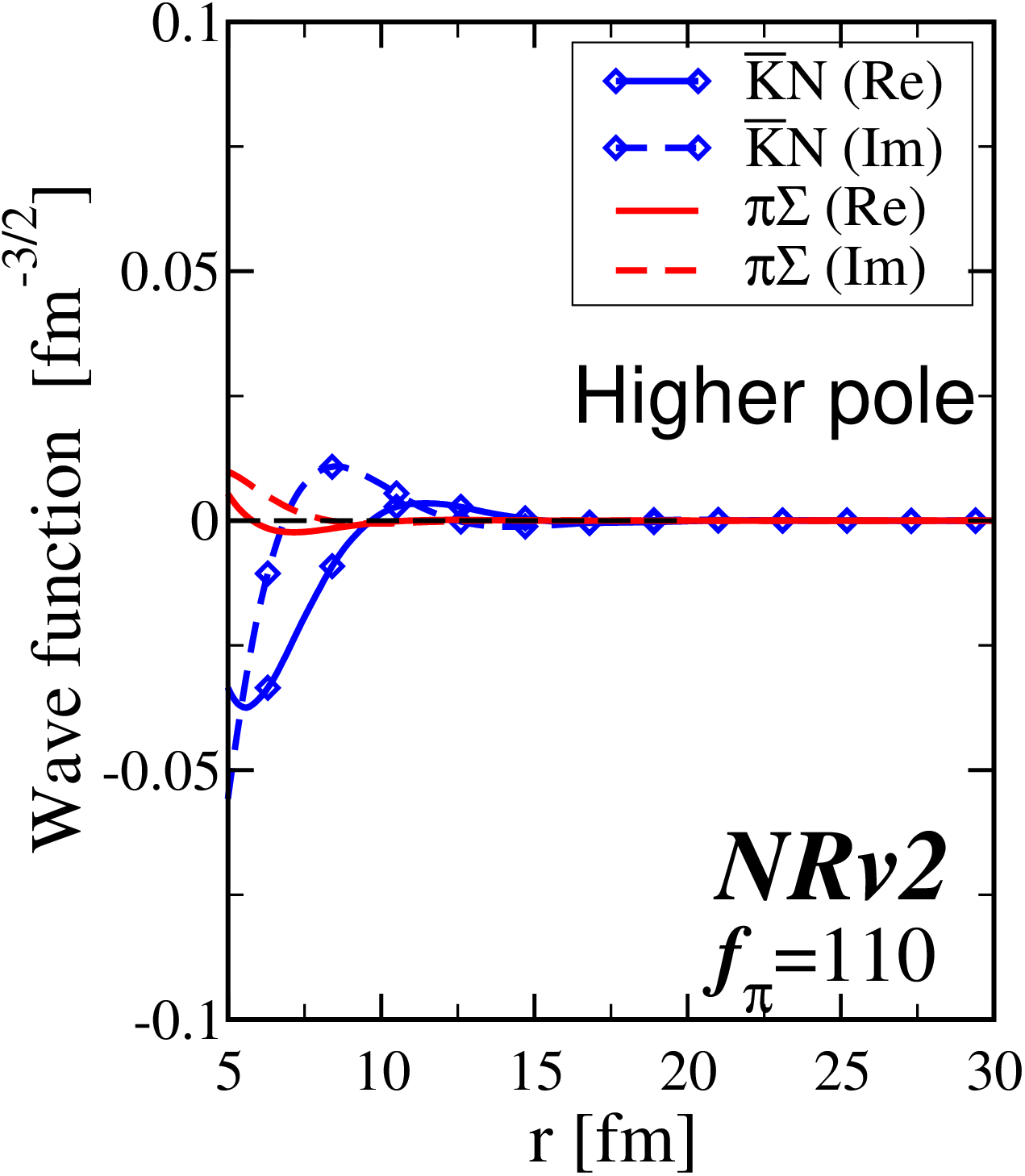}
  \caption{The ccCSM wave function of the higher-pole state 
of the NRv2 potential ($f_\pi=110$ MeV) 
under the same condition as the lower-pole wave function. 
Meaning of lines and symbols are the same as those 
of Fig. \ref{NRv2-Lower_wfnc}. 
 \label{NRv2-Higher_wfnc}}
\end{figure}

Figs. \ref{NRv2-Lower_wfnc} and \ref{NRv2-Higher_wfnc} show 
the ccCSM wave functions of the lower-pole and higher-pole 
states, respectively. In the calculation of those states, 
we use the complex-range Gaussian basis with $\omega=2.0$. 
The complex energy $(M, -\Gamma/2)$ of the pole considered here is 
$(1395.2, -137.9)$ MeV for the lower pole and $(1417.9, -16.6)$ MeV for the higher pole. 
As shown in Fig. \ref{NRv2-Lower_wfnc}, the $\pi\Sigma$ component dominates 
in the lower-pole wave function. The wave function, particularly at large distance, consists of almost 
only the $\pi\Sigma$ component, while the $\bar{K}N$ component exists at 
small distance. We can see an oscillatory behavior of the $\pi\Sigma$ 
wave function at far distance. Such the oscillatory behavior can be described 
owing to the complex-range Gaussian basis. 
On the other hand, the higher-pole wave function shown in 
Fig. \ref{NRv2-Higher_wfnc} is dominated by the $\bar{K}N$ component 
which is well localized at small distance as reported in the previous 
paper \cite{ccCSM-KN_NPA}. 
In the figure, we see almost no oscillatory behavior in 
the higher-pole wave function at large distance, in spite that 
the complex-range Gaussian basis is used for the calculation. 
We consider that the higher-pole state is almost equivalent to 
the $\bar{K}N$ bound state. 

\begin{figure}
  \includegraphics[width=.31\textwidth]{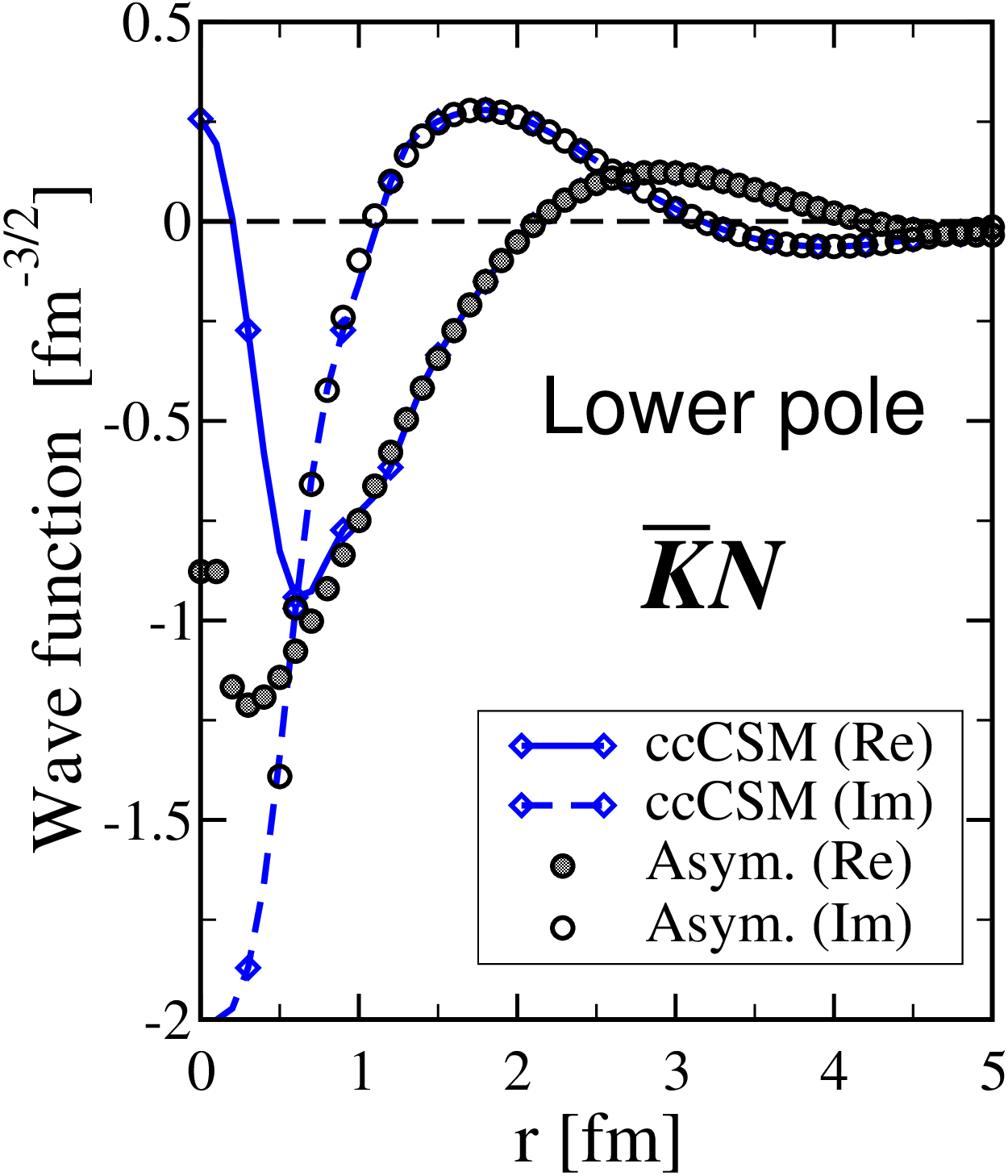}
  \hspace{0.0cm}
  \includegraphics[width=.32\textwidth]{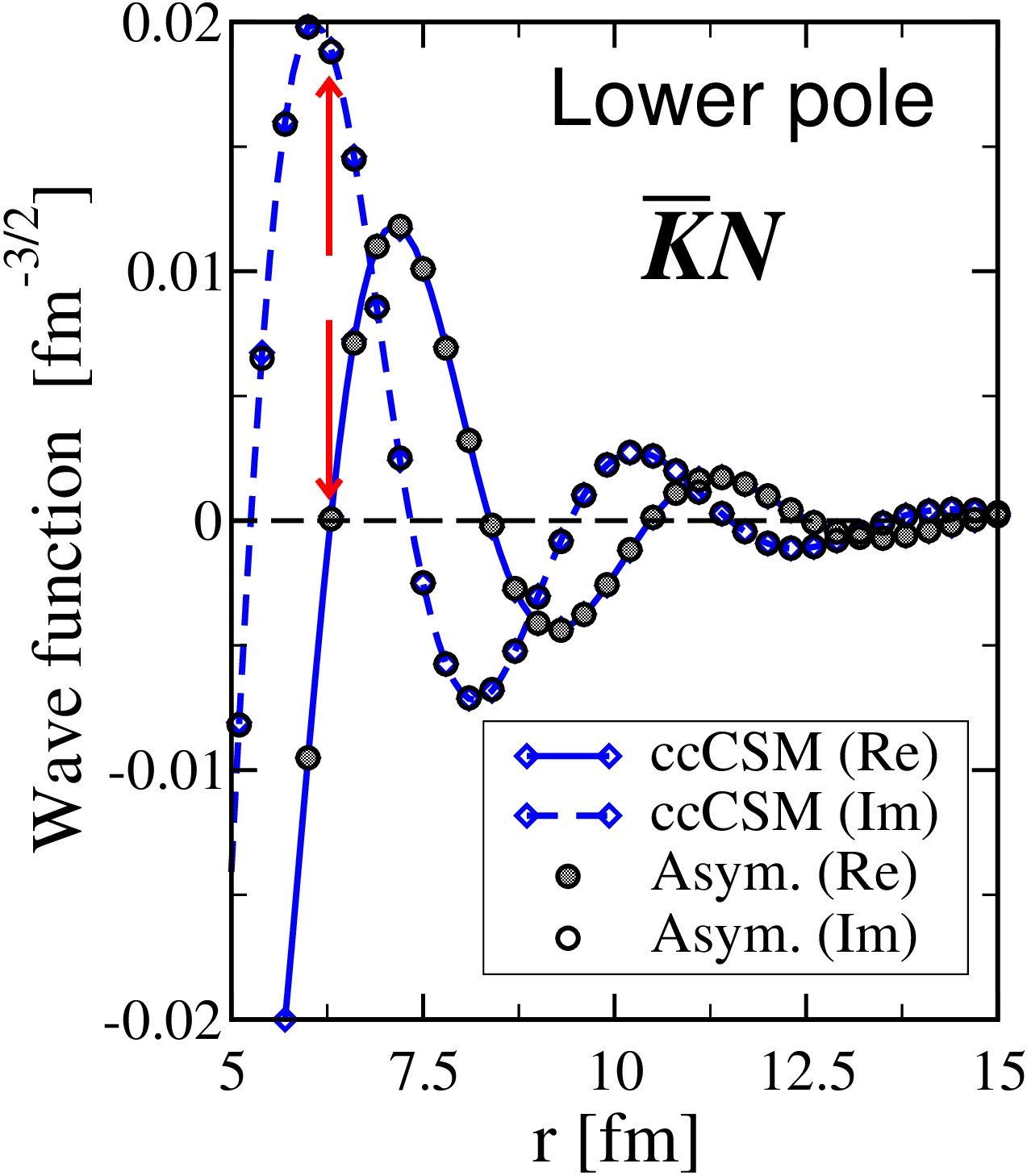}
  \hspace{0.0cm}
  \includegraphics[width=.32\textwidth]{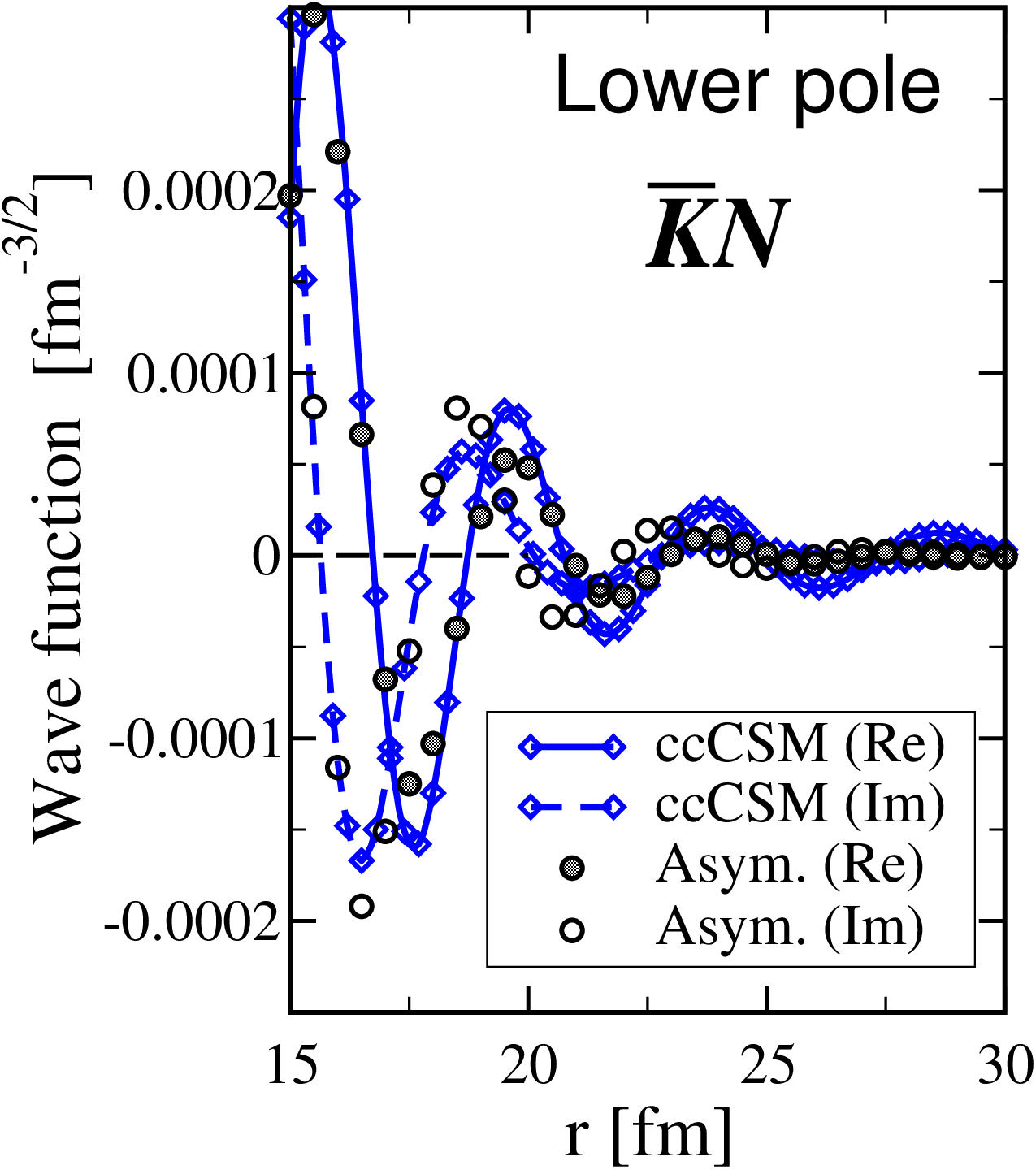}
  \caption{Comparison of the ccCSM wave function with the asymptotic 
wave function. ($\bar{K}N$ channel, Lower pole) 
The ccCSM wave function is that shown in Fig. \ref{NRv2-Lower_wfnc}. 
The real and imaginary parts of asymptotic wave function are depicted 
with black open and filled circle, respectively. 
The red arrow indicates the connection point at which the ccCSM 
wave function is connected to the asymptotic wave function. 
Left: $r=0\sim5$ fm, middle: $r=5\sim15$ fm, right: $r=15\sim30$ fm. 
\label{Lower_as_KN}}
\end{figure}

\begin{figure}
  \includegraphics[width=.31\textwidth]{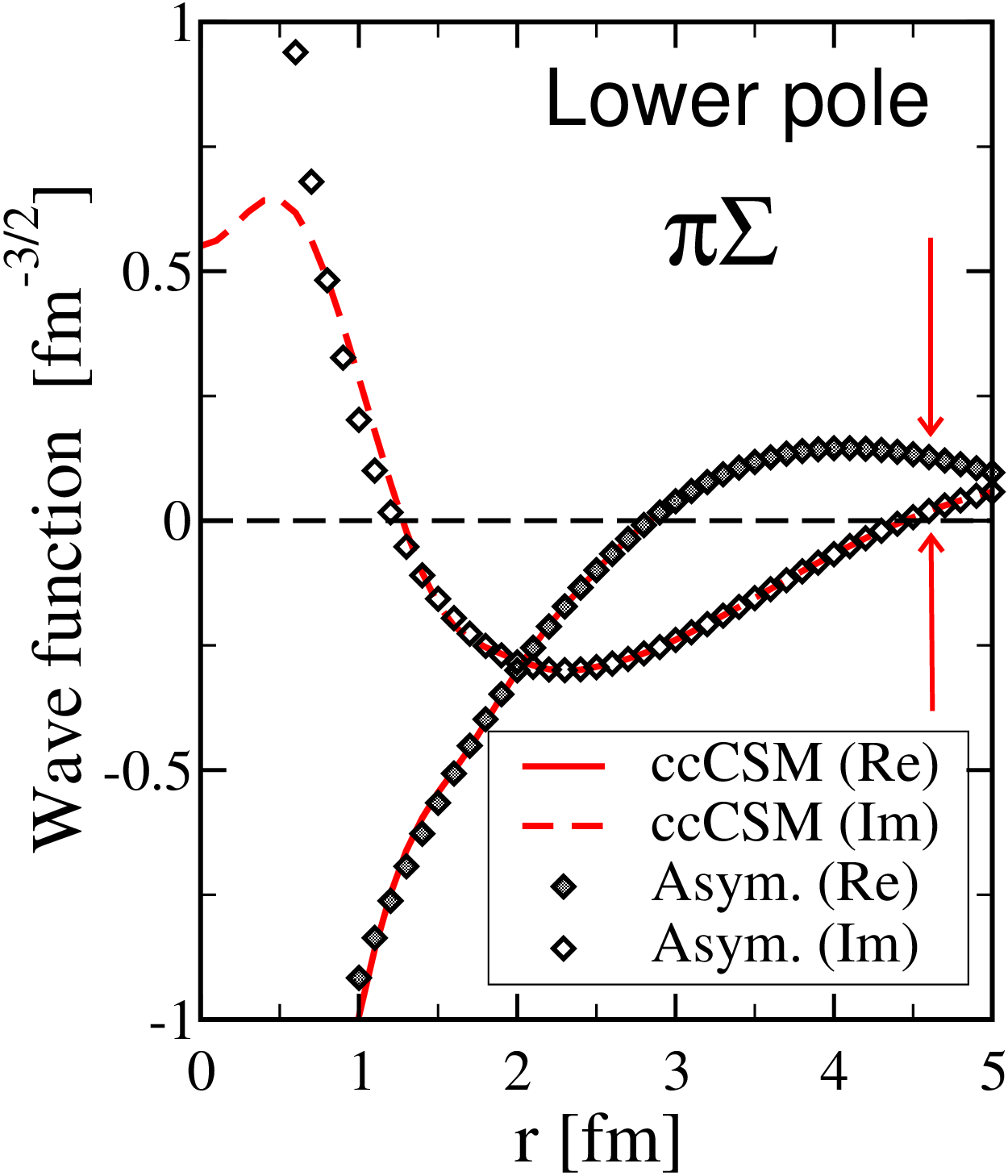}
  \hspace{0.0cm}
  \includegraphics[width=.32\textwidth]{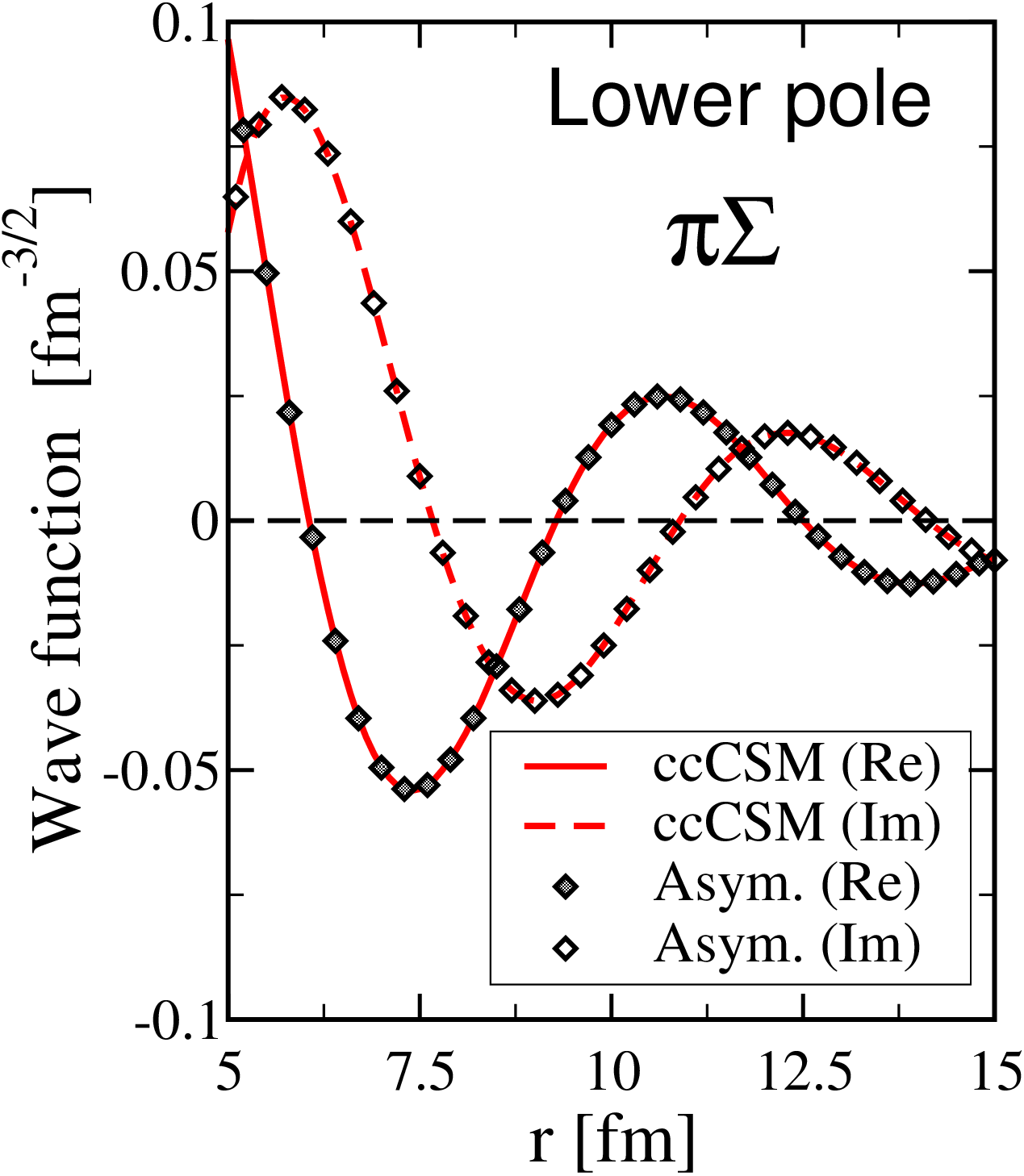}
  \hspace{0.0cm}
  \includegraphics[width=.32\textwidth]{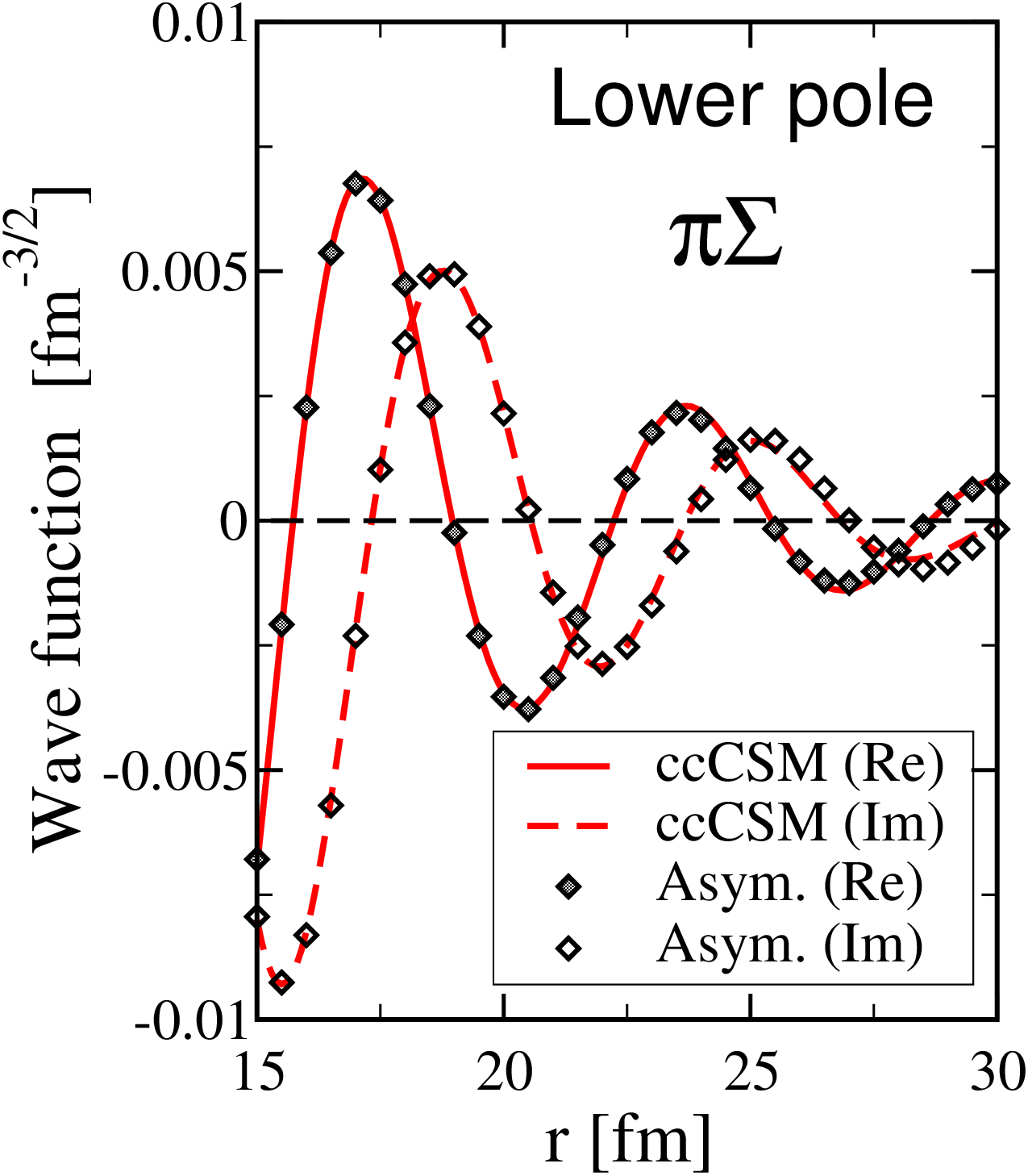}
  \caption{Comparison of the ccCSM wave function with the asymptotic 
wave function. ($\pi\Sigma$ channel, Lower pole) 
The asymptotic wave function is depicted with diamond symbols. 
Others are similar to Fig. \ref{Lower_as_KN}. 
 \label{Lower_as_pS}}
\end{figure}

Using the complex-range Gaussian basis, we have obtained a stable 
solution of the lower pole as a result of the improvement of 
the radial behavior of the resonance wave function. 
To check the reliability of the lower-pole wave function expanded with 
the complex-range Gaussian basis, we compare the ccCSM 
wave function with its asymptotic form which is determined analytically
by the complex resonance energy. 
The asymptotic form of a resonance wave function is known to be 
$\exp(i k_R r)/r$ with the complex wave number $k_R$. 
The complex-scaled asymptotic wave function is given as 
\begin{eqnarray}
\psi^{(c)}_{Asymp} (r_\theta) \; = \; 
A^{(c)} \, \frac{\exp(i k_R^{(c)} r_\theta)}{r_\theta}, \label{Asymp_wfunc}
\end{eqnarray}
where the symbol $(c)$ indicates a channel and $r_\theta$ is 
$r e^{i\theta}$ with 
the scaling angle $\theta$. The complex wave number $k_R^{(c)}$ in a channel 
$c$ is calculated with the lower-pole energy $Z_L$ as 
$\sqrt{2\mu_c (Z_{L}-M_c)}/\hbar$ where $M_c$ and $\mu_c$ are the total mass 
and reduced mass in the channel $c$, respectively. 
Here, we remark on the sign of the imaginary part of $k_R^{(c)}$. 
The sign of ${\rm Im} \, k_R^{(c)}$ has to be negative 
for open channels since the resonance pole is on an unphysical sheet 
of the complex energy plane in open channels. 
On the other hand, it should be positive for closed channels, since 
the pole is on a physical sheet in closed channels. 
In the present case that the pole exists above the $\pi\Sigma$ threshold 
and below the $\bar{K}N$ threshold, we choose the sign of 
${\rm Im} \, k_R^{(c)}$ 
as ${\rm Im} \, k_R^{(\bar{K}N)} > 0$ and ${\rm Im} \, k_R^{(\pi\Sigma)} < 0$. 
We determine the complex-valued coefficient $A^{(c)}$ and 
the connection point $s_0^{(c)}$ in each channel so that the asymptotic 
wave function $\psi^{(c)}_{Asymp} (r_\theta)$ connects 
most smoothly to the lower-pole wave function 
which is obtained by the ccCSM calculation. 
 
\begin{figure}
  \centering
  \includegraphics[width=.5\textwidth]{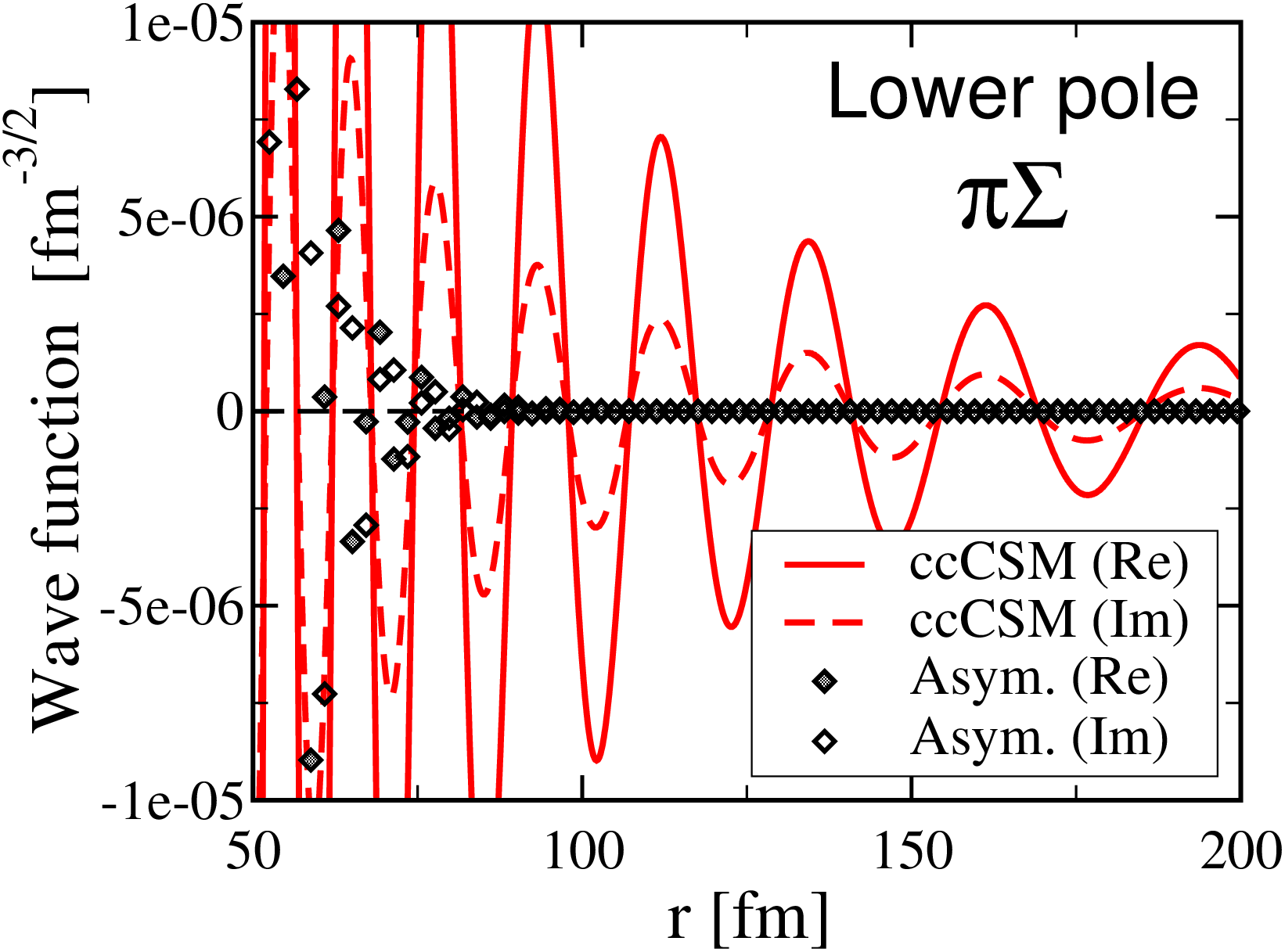}
  \caption{Comparison of the ccCSM wave function ($\pi\Sigma$ channel, Lower pole) with its asymptotic wave function at far distance ($r=50\sim200$ fm). 
\label{Lower_as_pS_Long}}
\end{figure}

In Figs. \ref{Lower_as_KN} and \ref{Lower_as_pS}, we show 
the $\bar{K}N$ and $\pi\Sigma$ components of the ccCSM wave functions 
together with those asymptotic forms given as Eq. (\ref{Asymp_wfunc}), 
respectively. The ccCSM wave functions of $\bar{K}N$ and $\pi\Sigma$ 
are connected to the asymptotic forms at $s_0^{(\bar{K}N)}=5.96$ fm 
and $s_0^{(\pi\Sigma)}=4.66$ fm, respectively. It is confirmed that 
the $\bar{K}N$ ($\pi\Sigma$) component of 
the ccCSM wave function well coincides with its asymptotic form 
upto $\sim 20$ ($30$) fm. 
In addition, the asymptotic wave function is found to reproduce 
the ccCSM wave function even in the short distance region ($\sim 1$ fm). 
It is worth to check how far in the distance the ccCSM wave function 
agrees with the asymptotic one. 
For this purpose, we show the $\pi \Sigma$ wave function 
in a range of $r=50\sim200$ fm in Fig. \ref{Lower_as_pS_Long}. 
The ccCSM wave function of the $\pi\Sigma$ component 
deviates from the asymptotic wave function at very large distance. 
It is noticed that the $\pi\Sigma$ wave function 
retains a very small but sizable magnitude even in more than 100 fm, 
whereas the asymptotic wave function is completely damped down there. 
This deviation does not affect the normalization of the wave function 
because of the relatively very small amplitudes at this large distance region. 

Thus, the complex-range Gaussian basis is found to be effective 
for the study of the double-pole structure of $\Lambda(1405)$. 
By using the complex-range Gaussian basis, the lower pole as well as 
the higher pole is obtained as a stable eigen solution. 
The ccCSM wave functions of both poles are confirmed 
to reproduce well the asymptotic behavior of resonance wave function 
within the distance of around 30 fm. 
Hereafter, we perform further analysis of the double-pole states 
with these wave functions. 

\begin{table}
\caption{Composition of the lower- and higher-pole states at 
$f_\pi=90$ to 120 MeV. 
Both states are calculated with the complex-range Gaussian basis. 
``Norm ($\bar{K}N$)'' and ``Norm ($\pi\Sigma$)'' indicate 
norms of the $\bar{K}N$ and $\pi\Sigma$ components, respectively. 
Detailed explanation is given in the text. The potential NRv2 is used. 
\label{Norm-NRv2}}
\begin{tabular*}{\textwidth}{l@{\hspace{0.7cm}}|@{\hspace{0.7cm}}r@{\hspace{1.0cm}}r@{\hspace{1.0cm}}r@{\hspace{1.0cm}}r}
\hline
 $f_\pi$ & 90 & 100 & 110 & 120 \\
\hline
Lower pole&&&& \\
 Norm ($\bar{K}N$)  & $-0.076$  & $-0.051$  & $0.097$   &  $0.136$ \\
                    & $+0.022i$ & $+0.225i$ & $+0.154i$ &  $+0.120i$ \\
 Norm ($\pi\Sigma$) & $1.076$   & $1.051$   & $0.903$   & $0.864$ \\
                    & $-0.022i$ & $-0.225i$ & $-0.154i$ & $-0.120i$ \\
\hline
Higher pole&&&& \\
 Norm ($\bar{K}N$)  & $0.960$ & $1.066$ & $1.115$ & $1.120$ \\
                    & $+i0.174$ & $+0.160i$ & $+0.098i$ & $+0.044i$ \\
 Norm ($\pi\Sigma$) & $0.040$ & $-0.066$ & $-0.115$ & $-0.120$ \\
                    & $-i0.174$ & $-0.160i$ & $-0.098i$ & $-0.044i$ \\
\hline
\end{tabular*}
\end{table}

Table \ref{Norm-NRv2} shows the composition 
of the higher- and lower-pole states. 
The norm of each component is calculated as 
$\langle \tilde{\phi}^{l=0, \theta}_\alpha | 
\phi^{l=0, \theta}_\alpha \rangle$ ($\alpha=\bar{K}N, \, \pi\Sigma$), 
where $\phi^{l=0, \theta}_\alpha (r)$ is the spatial part of 
the ccCSM wave function in the channel $\alpha$, defined in Eq. (\ref{Eq:GaussE_CSM}). 
We expect that this quantity is useful to know the composition of 
the resonances, while the quantity is complex value. 
It is noted that the sum of the quantities ``Norm ($\bar{K}N$)'' 
and ``Norm ($\pi\Sigma$)'' in Table \ref{Norm-NRv2} 
should be unity due to the normalization of the state, as explained 
in the section \ref{Formalism}. 
For the lower pole, the $\pi\Sigma$ norm is apparently larger 
magnitude than the $\bar{K}N$ norm, wheres the $\bar{K}N$ norm is reversely 
large magnitude for the higher pole. This result also means that the major 
component is the $\pi\Sigma$ in the lower pole, and 
the $\bar{K}N$ in the higher pole, as Figs. \ref{NRv2-Lower_wfnc} and \ref{NRv2-Higher_wfnc} indicated. 

\begin{table}
\caption{Meson-baryon mean distance calculated with the ccCSM wave function 
connected to its asymptotic form. 
the lower- and higher-pole states given in 
Table \ref{Norm-NRv2} are considered. 
The values of $\sqrt{\langle r^2 \rangle}_{\bar{K}N}$, $\sqrt{\langle r^2 \rangle}_{\pi\Sigma}$ and $\sqrt{\langle r^2 \rangle}_{\bar{K}N+\pi\Sigma}$ are the mean distance 
of the $\bar{K}N$ component, the $\pi\Sigma$ component and total system, respectively. All values are given in unit of fm. 
\label{Distance-NRv2_asymp}}
\begin{tabular*}{\textwidth}{l|@{\hspace{0.5cm}}rrrr}
\hline
 $f_\pi$ & 90 & 100 & 110 & 120 \\
\hline
Lower pole&&&& \\
 $\sqrt{\langle r^2 \rangle}_{\bar{K}N}$  
& $0.87-0.13i$ & $0.75-0.18i$ & $0.66-0.20i$ & $0.48-0.03i$ \\
 $\sqrt{\langle r^2 \rangle}_{\pi\Sigma}$ 
& $0.09-0.65i$ & $0.65-0.13i$ & $0.04-0.51i$ & $0.01+0.48i$ \\
 $\sqrt{\langle r^2 \rangle}_{\bar{K}N+\pi\Sigma}$ 
& $0.13-0.43i$ & $0.67-0.18i$ & $0.06-0.46i$ & $0.08+0.41i$ \\
\hline
Higher pole&&&& \\
 $\sqrt{\langle r^2 \rangle}_{\bar{K}N}$ 
& $1.22-0.47i$ & $1.25-0.42i$ & $1.28-0.40i$ & $1.31-0.39i$ \\
 $\sqrt{\langle r^2 \rangle}_{\pi\Sigma}$ 
& $0.29+0.92i$ & $0.26+0.92i$ & $0.23+0.93i$ & $0.21+0.93i$ \\
 $\sqrt{\langle r^2 \rangle}_{\bar{K}N+\pi\Sigma}$ 
& $1.27-0.24i$ & $1.40-0.26i$ & $1.47-0.33i$ & $1.49-0.38i$ \\
\hline
\end{tabular*}
\end{table}

We further consider the size of the lower-pole state, as we did for 
the higher-pole state in the previous paper \cite{ccCSM-KN_NPA}. 
Certainly, the lower-pole wave function is well described with 
the complex-range Gaussian basis. 
However, the description at far distance is found to be insufficient:  
the ccCSM wave function of the lower-pole state 
deviates from the asymptotic form, as shown in Fig. \ref{Lower_as_pS_Long}. 
Although the amplitude of the ccCSM wave function is very small to 
be $\sim 10^{-5}$ at 50 fm, this tiny deviation 
causes a problem in the calculation of 
the mean distance between the meson and the baryon ($MB$ distance);  
the convergence of $MB$ distance is very slow with respect to 
the basis number $N$, 
in spite that the pole energy has been sufficiently converged.
To overcome this problem, 
we utilize the asymptotic form of the ccCSM wave function, Eq. (\ref{Asymp_wfunc}), {\it only} for the calculation of the $MB$ distance. 
We use the ccCSM wave function in a short region $r<s_0^{(c)}$ of 
the connection point and the asymptotic wave function 
outside the connection point ($r>s_0^{(c)}$). 
Calculated in this way at each $f_\pi$ value given in the Hamiltonian, 
the mean distance of the lower-pole state 
is stable for the number of basis as well as the scaling angle. 
The result is shown in Table \ref{Distance-NRv2_asymp}. 
It is found that the mean distance is  
extremely a small value for the real part, especially 
in the $\pi\Sigma$ component. 
We consider that in such broad resonances like 
the lower-pole state, it is difficult 
to introduce the physical meaning of the expectation value 
of the mean distance. 
As shown in Table \ref{Distance-NRv2_asymp}, 
the mean distance for the lower-pole state has 
large imaginary part, which is comparable to the real part.  
This situation is different from the case of the higher-pole state in which 
the imaginary part of energy and mean distance is still small. 
Hence we can not make an intuitive interpretation of the calculated 
mean distance in the lower-pole case, 
even though its matrix element is stably obtained for the Hamiltonian 
with a $f_\pi$ value given. 
It is noted that 
the mean distance of the higher-pole state given in the table 
is not so different from that reported in our previous 
paper \cite{ccCSM-KN_NPA} 
in which the asymptotic form is not used. 

At the end of this section, we comment on 
interpretation of the $MB$ distance obtained in the ccCSM calculation. 
In the ccCSM, a resonant state is treated as a Gamow state. 
The $MB$ distance is calculated as a expectation 
value of $\hat{\bf r}^2$ with the wave function of the Gamow state. 
($\hat{\bf r}$ is the operator of 
the relative coordinate.) 
On the other hand, such the $MB$ distance can be extracted from 
the electro-magnetic (EM) form factor of $\Lambda(1405)$ when 
it is treated as a meson-baryon molecular state. 
According to Ref. \cite{ChU:EMsize} which is 
based on chiral dynamics as well as our study, it is found that  
the $MB$ distance of the higher pole obtained through the EM form factor is 
almost identical to that obtained in the ccCSM. It is noted that 
the higher-pole energy is almost the same between both studies. 
Therefore, we consider that 
the $MB$ distance in the ccCSM is related to the form factor 
of the resonance.

\subsection{Double-pole nature and the $\bar{K}N$-$\pi\Sigma$ coupling potential}
\label{}

\begin{figure}
  \centering
  \includegraphics[width=.55\textwidth]{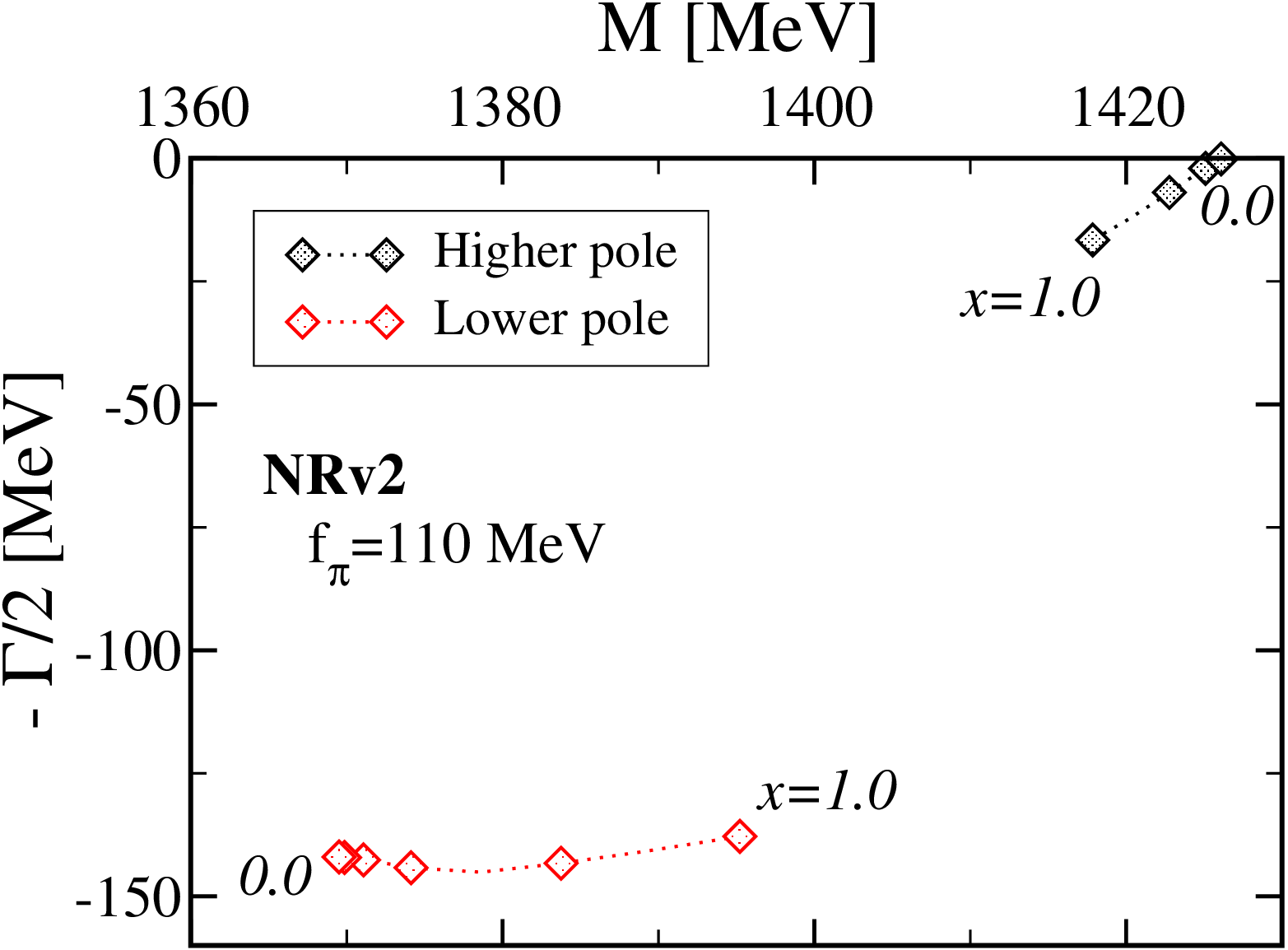}
\caption{Behavior of the higher and lower poles on the complex-energy plane 
when the strength of 
the $\bar{K}N$-$\pi\Sigma$ coupling potential ``$x$'' is modified. 
NRv2 potential with $f_\pi=110$ MeV is considered. 
All poles are obtained with $\theta=40^\circ$. 
The $\pi\Sigma$ and $\bar{K}N$ thresholds are located at 1331 MeV and 1435 MeV, respectively. 
\label{EGVs_DP-NRv2-fp=110_Cmod}}
\end{figure}

\begin{figure}
\centering
  \includegraphics[width=.33\textwidth]{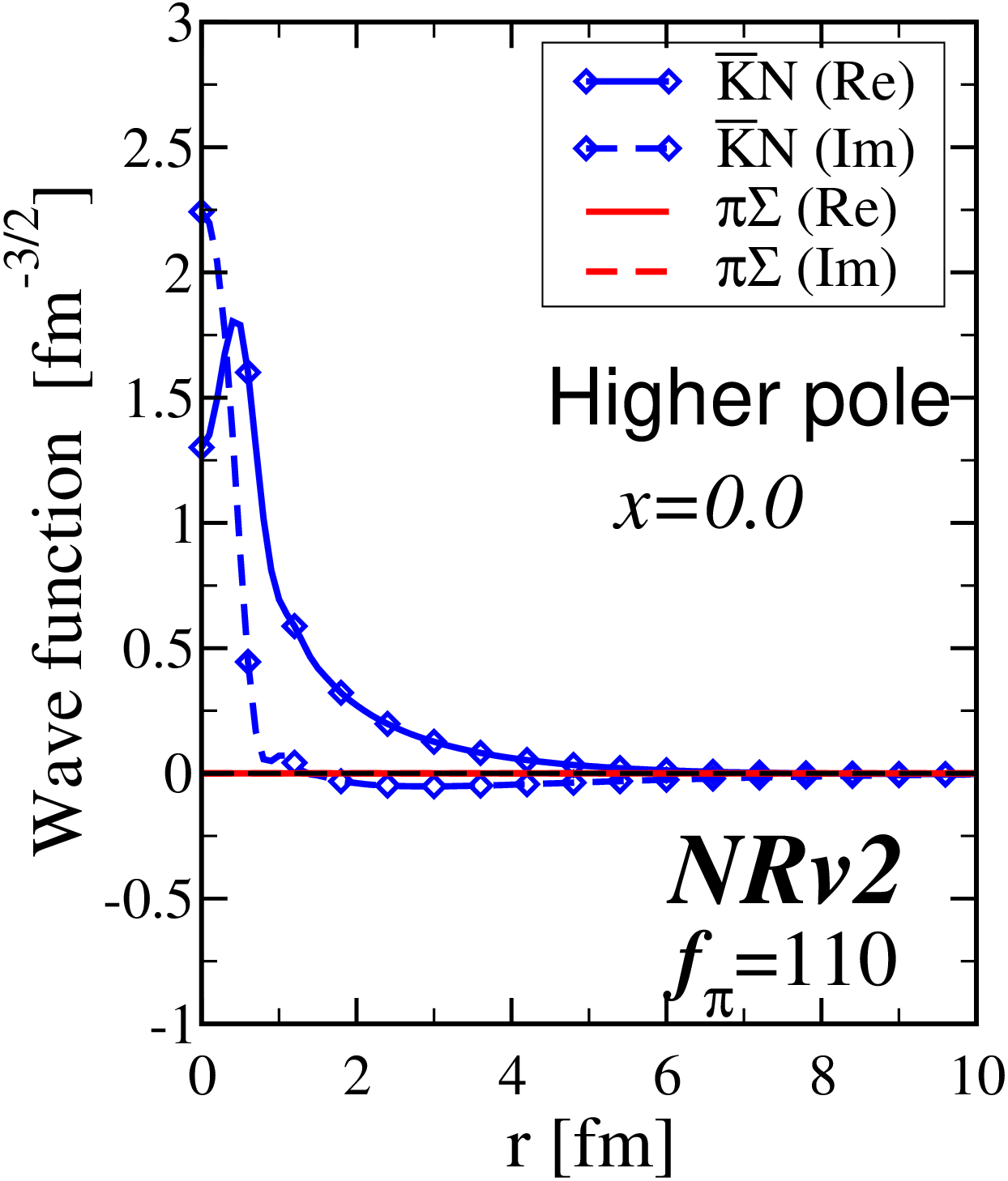}
  \hspace{1.0cm}
  \includegraphics[width=.33\textwidth]{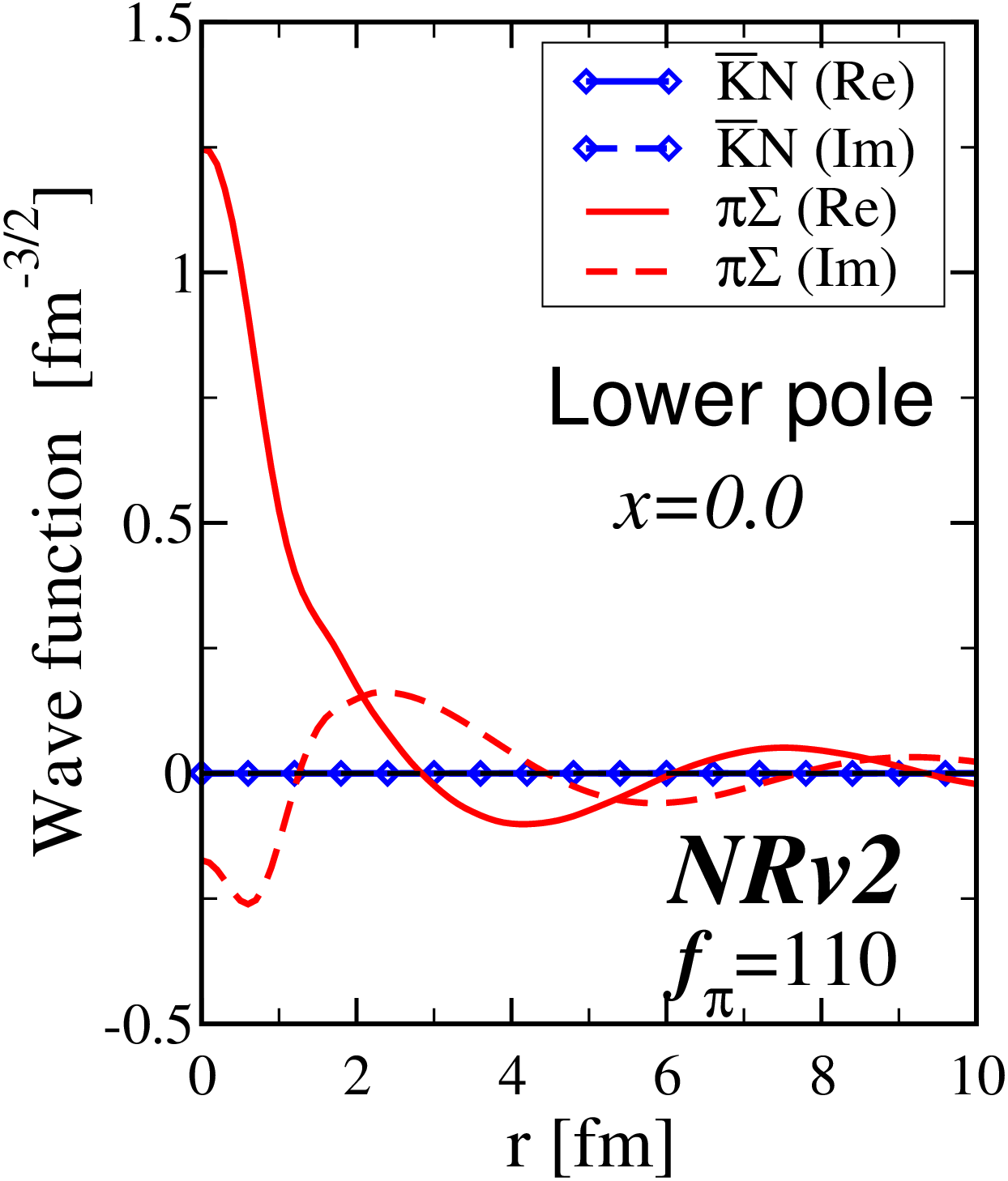}
  \caption{The ccCSM wave function of the double pole
of the NRv2 potential in which the $\bar{K}N$-$\pi\Sigma$ coupling potential 
is switched off ($x=0.0$). Left (right) panel shows the higher-pole 
(lower-pole) wave function. $f_\pi=110$ MeV and $\theta=40^\circ$. 
Meaning of lines and symbols are the same as those of Fig. \ref{NRv2-Lower_wfnc}. 
\label{NRv2-HL-x=0_wfnc}}
\end{figure}

To investigate the double-pole structure in detail, 
we have examined to vary the strength of the $\bar{K}N$-$\pi\Sigma$ 
coupling potential; 
\begin{equation}
V^{(I=0)}_{\bar{K}N,\pi\Sigma} (r) \quad 
\rightarrow \quad x \; V^{(I=0)}_{\bar{K}N,\pi\Sigma} (r), 
\end{equation}
where a real parameter $x$ is introduced to scale the strength of 
the coupling potential. 
Note that a similar analysis has been performed with a chirally motivated 
model \cite{ChMM:CS-pole_move}. 
Fig. \ref{EGVs_DP-NRv2-fp=110_Cmod} shows the behavior of higher and 
lower poles when $x$ is reduced from 1.0 to 0.0, namely no coupling limit. 
For each value of $x$, we have obtained both of higher and lower 
poles, taking into 
account the self-consistency for the complex energy as done so far. 
As shown in the figure, with the parameter $x$ decreasing, 
the higher pole approaches to the real energy axis. 
At $x=0$, it comes onto the real energy axis below the $\bar{K}N$ threshold located at 1435 MeV. 
The higher pole becomes a $\bar{K}N$ bound 
state in case of no $\bar{K}N$-$\pi\Sigma$ coupling potential. 
The $\bar{K}N$ binding energy is 8.8 MeV and the mean distance is 2.08 fm. 
On the other hand, the lower pole moves to 
a smaller mass region with keeping the imaginary energy. 
Even at the no-coupling limit ($x=0$), the lower pole remains to be 
a resonance. 
We mention to earlier studies. A prior research based on the chiral 
unitary model within $\bar{K}N$ and $\pi\Sigma$ channels \cite{ChU:HW} has 
shown qualitatively a similar result at the no coupling limit. 
The same conclusion on the double pole has been reported also by a study 
with a chirally motivated model \cite{ChMM:CS-pole_move} which involves 
more channels and utilizes the latest experimental data of kaonic 
hydrogen atom \cite{KpX:SIDDHARTA}.

We show ccCSM wave functions of both poles 
without the channel coupling at $x=0$ in Fig. \ref{NRv2-HL-x=0_wfnc}. 
The higher-pole state purely consists of the $\bar{K}N$ 
component, whereas the lower-pole state contains only the $\pi\Sigma$ 
component. The $\bar{K}N$ wave function of the higher-pole state 
rapidly damps down, since this state is a $\bar{K}N$ bound state. 
In the $\pi\Sigma$ wave function of the lower pole, 
the oscillatory behavior continues in a distance, though 
the wave function damps due to the complex scaling. 
We comment on the higher pole. 
As mentioned, it becomes a $\bar{K}N$ bound state at the limit
of no $\bar{K}N$-$\pi\Sigma$ coupling. 
Once the coupling potential is switched on, this $\bar{K}N$ bound state 
couples to the $\pi\Sigma$ state. As a result, the $\bar{K}N$ bound state 
acquires a $\pi\Sigma$ decay width to be a resonance. 
Therefore, the higher-pole state can be understood to 
be a Feshbach resonance. 

\begin{table}
\caption{Norm of the double pole at each strength of 
the coupling potential ($x$). (NRv2 potential with $f_\pi=110$ MeV. 
$\theta=40^\circ$) 
``Norm ($\bar{K}N$)'' indicates the norm of $\bar{K}N$ component in each 
state.
\label{Norm-NRv2_Cmod}}
\begin{tabular*}{\textwidth}{l@{\hspace{0.3cm}}|@{\hspace{0.3cm}}r@{\hspace{0.8cm}}r@{\hspace{0.8cm}}r@{\hspace{0.8cm}}r@{\hspace{0.8cm}}r}
\hline
 $x$ & 0.0 & 0.3 & 0.5 & 0.8 & 1.0\\
\hline
Lower pole&&&&& \\
 Norm ($\bar{K}N$) & $0.000$   & $ 0.001$  & $ 0.007$  & $0.055$   & $0.097$\\
                   & $+0.000i$ & $+0.003i$ & $+0.011i$ & $+0.083i$ & $+0.154i$\\
\hline
Higher pole&&&&& \\
 Norm ($\bar{K}N$) & $1.000$   & $1.006$   & $1.018$   & $1.057$   & $1.115$\\
                   & $+0.000i$ & $+0.003i$ & $+0.010i$ & $+0.042i$ & $+0.098i$\\
\hline
\end{tabular*}
\end{table}
 
We have investigated how the property of the higher- and 
lower-pole states changes in scaling the coupling strength $x$. 
Table \ref{Norm-NRv2_Cmod} shows the norm of $\bar{K}N$ component 
in both states at each $x$. 
The composition is found to change continuously with the parameter $x$, 
as well as the poles move smoothly on the complex energy plane 
as shown in Fig. \ref{EGVs_DP-NRv2-fp=110_Cmod}. 
With the coupling strength $x$ enhanced, the magnitude of 
the $\bar{K}N$ norm increases from zero in the lower pole. 
The lower pole is purely the $\pi\Sigma$ state at the no coupling limit, 
since the $\bar{K}N$ norm is zero. The $\bar{K}N$ component is mixed 
gradually to the state, increasing the coupling strength. 
In the higher pole, the $\bar{K}N$ component is dominant,
 since the $\bar{K}N$ norm is close to unity.\footnote{Some norms of 
components exceed the unity, though the wave function is surely 
normalized to unity. Such a behavior of norms in the CSM has been 
shown also in the study of resonances of unstable nuclei \cite{CSM:Masui}.} 
Due to the normalization of the total wave function, 
the $\pi\Sigma$ norm can be calculated as $1-{\rm Norm(}\bar{K}N{\rm )}$. 
Hence, the $\pi\Sigma$ component is confirmed to be mixed 
in the higher pole, when the coupling strength increases.

\subsection{Double-pole nature and the energy dependence of potential}
\label{}

The most important point of the chiral SU(3)-based potential 
is its energy dependence due to the chiral dynamics. 
In this section, we consider the influence of the energy-dependence 
of the chiral SU(3)-based potential to the double-pole nature of 
the $I=0$ $\bar{K}N$-$\pi\Sigma$ system. Motivated by the earlier 
work \cite{KN-pS:IHJ}, we introduce an energy-independent version 
of our potentials.  

We make a consideration on this issue with the two non-relativistic 
versions of our potentials (NRv1 and 2). 
The essential structure of these potentials are 
\begin{equation}
V_{ij}(r) = -\frac{C_{ij}}{8f_\pi^2} (\omega_i+\omega_j) \times [\rm{flux \; factor}] \times (\rm{Gaussian \; form \; factor}), 
\end{equation}
where $\omega_i$ indicates the meson energy in the channel $i$. 
Thus, the NRv1 and 2 potentials are proportional to the meson energies 
$\omega_i+\omega_j$ which are attributed to the chiral dynamics. 
Here, we examine to remove this energy dependence by using 
static approximation for mesons in which the meson energies are replaced 
with meson masses $m_i + m_j$; $\omega_i+\omega_j \rightarrow 
m_i + m_j$.  
We denote the meson-energy-fixed potentials as ``NRv1(2) (E-indep)''. 
It should be noted that ``NRv2 (E-indep)'' potential 
is completely an energy-independent potential,
since the flux factor involved in the NRv2 potential has already lost 
the energy dependence by an approximation from original 
relativistic form to a non-relativistic one. 
On the other hand, the ``NRv1 (E-indep)'' potential has still 
an energy dependence such as $1/\sqrt{s}$ involved in the flux factor. 
(See Eqs. (7) and (8) in Ref. \cite{ccCSM-KN_NPA}.) 
Similarly to the original potentials, the range parameters 
of NRv1(2) (E-indep) are determined so as to reproduce the $I=0$ $\bar{K}N$ scattering length obtained by Martin's analysis.
\footnote{
In case of  $f_\pi=110$ MeV, the range parameters ($d_{\bar{K}N, \bar{K}N}$, $d_{\pi\Sigma, \pi\Sigma}$) are (0.459, 0.426) fm in the NRv1 (E-indep) and 
(0.457, 0.457) fm in the NRv2 (E-indep). They give 
the scattering length $a_{\bar{K}N(I=0)} = -1.701+0.677i$ fm and 
$-1.701+0.681i$ fm, respectively. 
Martin's value is $a_{\bar{K}N(I=0)} = -1.70+0.68i$ fm.}

\begin{figure}
  \includegraphics[width=.47\textwidth]{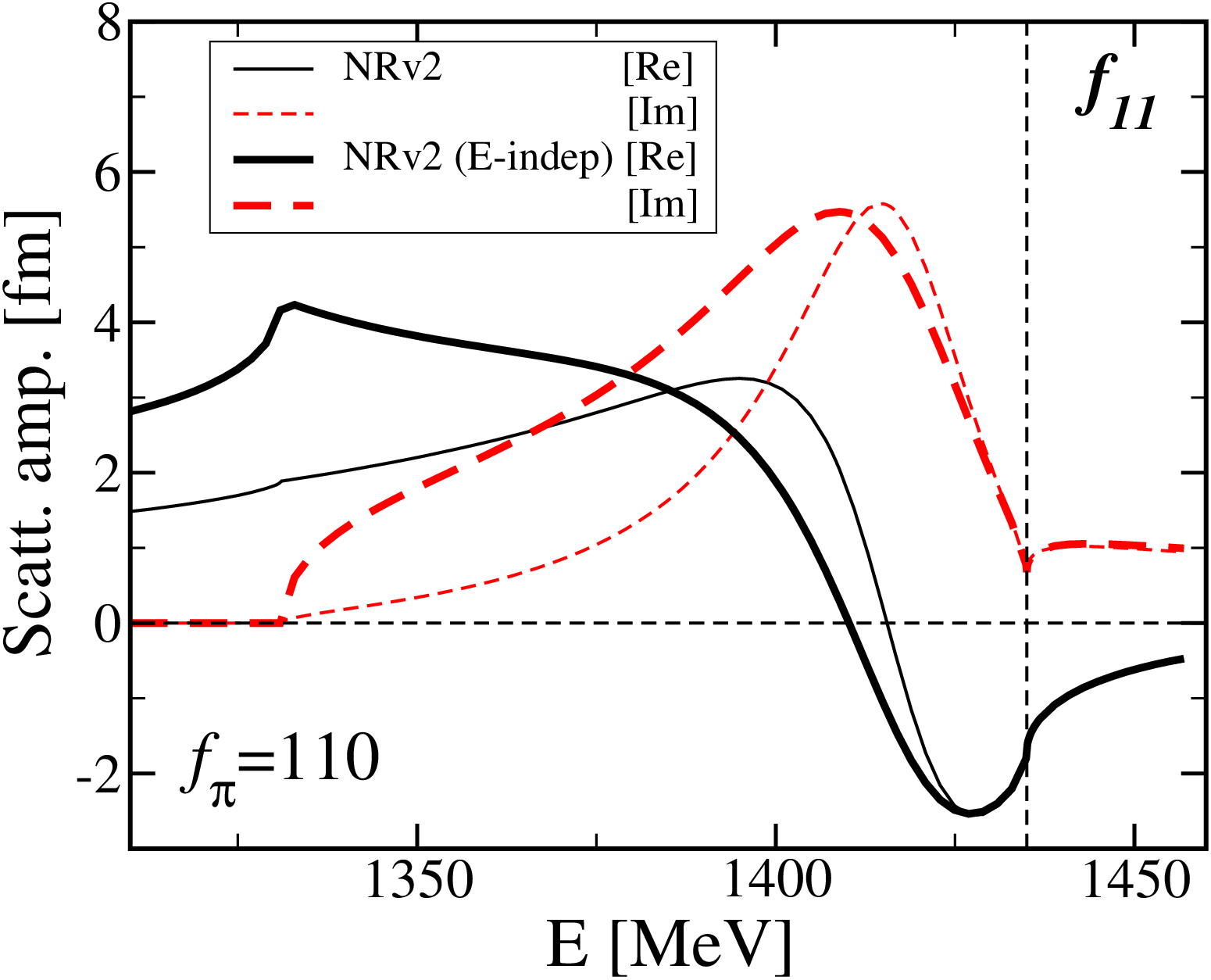}
  \hspace{0.3cm}
  \includegraphics[width=.47\textwidth]{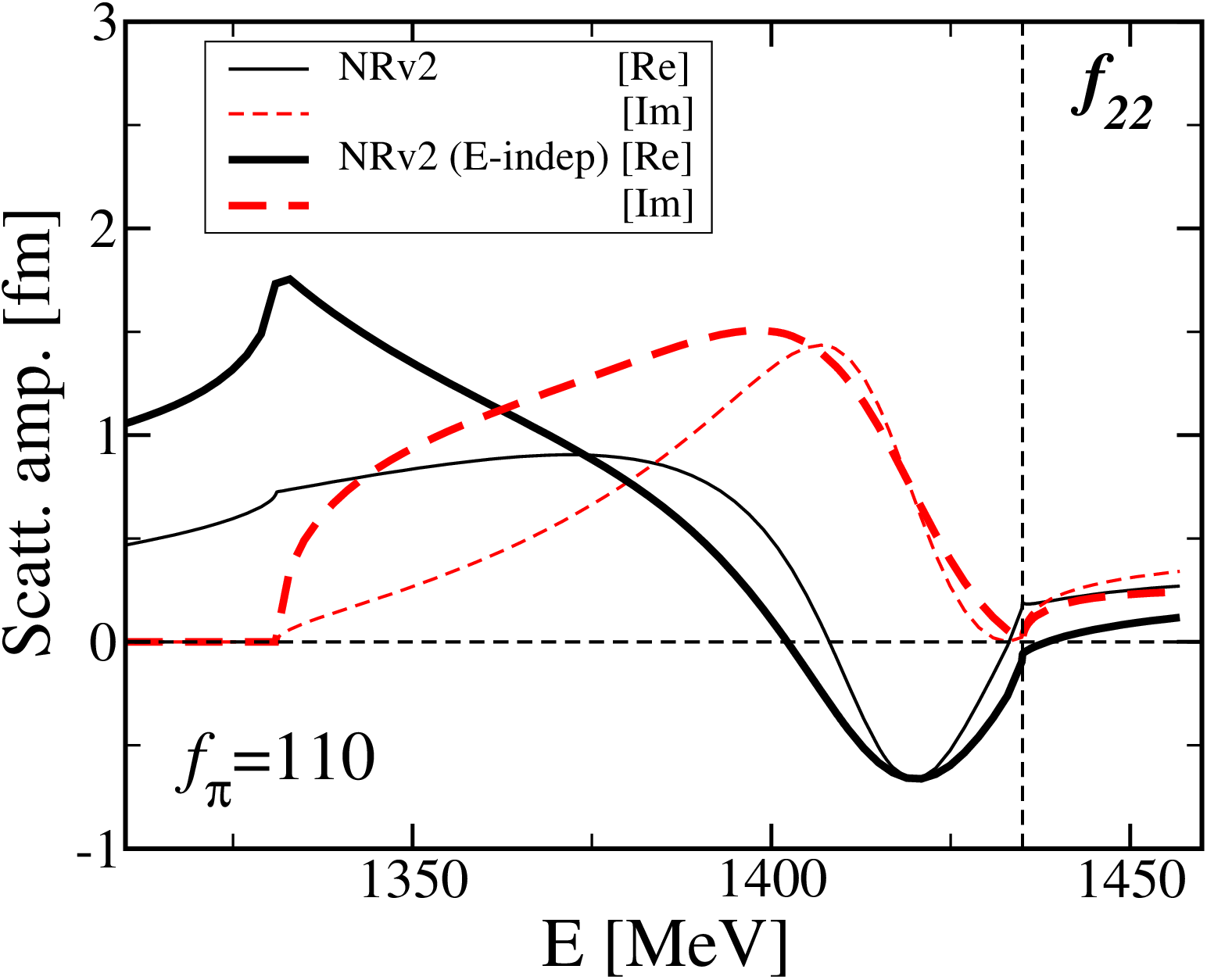}
  \caption{
Comparison of $I=0$ scattering amplitudes between 
the energy-independent case ``NRv2(E-indep.)'' and the energy-dependent case ``NRv2''.  ($f_\pi=110$ MeV) 
The scattering amplitudes of NRv2(E-indep.) and NRv2 are shown 
with bold and thin lines, respectively. 
The real (imaginary) part of scattering amplitude is drawn with 
a black-solid (red-dashed) line. 
Left (right) panel shows $\bar{K}N$ ($\pi\Sigma$) scattering amplitude. 
Vertical dashed line means the $\bar{K}N$ threshold. 
\label{ScattAmp_NRv2s}}
\end{figure}

We investigate the difference between the energy-dependent 
and energy-independent potentials in the scattering amplitude. 
Fig. \ref{ScattAmp_NRv2s} shows the $\bar{K}N$ 
and $\pi\Sigma$ amplitudes calculated with NRv2 and NRv2 (E-indep) potentials. 
As seen in the figure, both potentials give the similar 
scattering amplitudes near the $\bar{K}N$ threshold. However, below the resonance the scattering amplitudes are very different between two potentials. 
The energy-independent case (NRv2 (E-indep)) shows very attractive behavior 
in the low energy region, compared to the energy-dependent case (NRv2). 
The earlier work \cite{KN-pS:IHJ} has reported a similar result 
that the $\pi\Sigma$ attraction enhances in the energy-independent case. 
In their analysis, the $\pi\Sigma$ scattering length is found to become 
more attractive compared to the energy-dependent case.

\begin{figure}
  \centering
  \includegraphics[width=.55\textwidth]{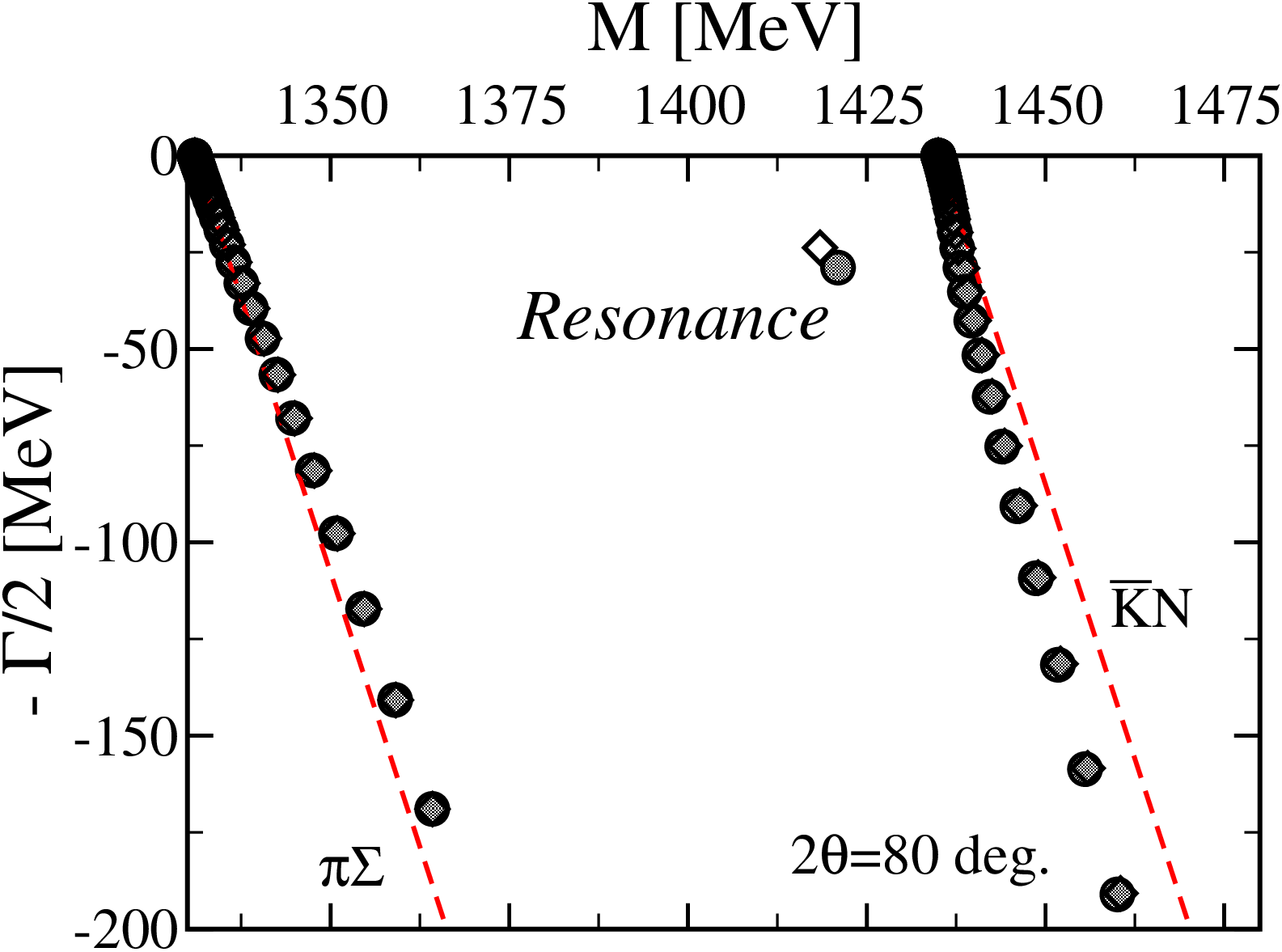}
\caption{Distribution of complex eigenvalues calculated with 
an energy-independent version of NRv1 and 2 potentials ($f_\pi=110$ MeV). 
Filled circles and open diamonds represent eigenvalues calculated 
with NRv1 and 2 potentials, respectively. 
The scaling angle $\theta$ is set to be $40^\circ$. 
Dashed lines are $2\theta$ lines indicating the $\bar{K}N$ and 
$\pi\Sigma$ continuum states. 
\label{EGVs_Eindep_fp=110}}
\end{figure}

We report resonance poles of the energy-independent potentials. 
Fig. \ref{EGVs_Eindep_fp=110} shows the complex eigenvalues 
of the energy-independent potentials (NRv1 and 2 (E-indep)). 
The scaling angle $\theta$ is set to be $40^\circ$. 
In Fig. \ref{EGVs_Eindep_fp=110}, 
only one eigenvalue is apparently apart from $2\theta$ lines in each potential 
and it can be identified to a resonance pole. 
As summarized in Table \ref{PolePosition-fp=110}, the pole position of this resonance ($Z_H$ of E-indep.) is close to that of the higher pole 
of the energy-dependent case ($Z_H$ of E-dep.). 
The real energy ($M$; mass) of the higher pole is 
almost the same between E-dep. and E-indep. potentials. 
Its imaginary energy ($\Gamma/2$; half decay width) tends to be larger 
 in the E-indep. potential than in the E-dep. potential. 
No other poles are found 
around the lower-pole region. Similar results are obtained for other $f_\pi$ 
values. In the NRv1 (E-indep) case, we searched for self-consistent solutions 
with respect to the pole energy since the energy dependence still remains in 
the flux factor. There are no self-consistent solutions 
corresponding to the lower pole when we scanned such the region as 
$M_N+m_K-100{\rm MeV} \leqslant {\rm Re} \, Z \leqslant M_N+m_K$ and 
$-200 \leqslant {\rm Im} \, Z \leqslant -50$ MeV on the complex-energy plane. 
($M_N+m_K$ means the $\bar{K}N$ threshold.) 
Therefore, the double-pole structure of 
$I=0$ $\bar{K}N$-$\pi\Sigma$ system is attributed to the energy dependence 
of the chiral SU(3) potential. We consider that the lower pole, in particular, 
is generated by this energy dependence. 

It should be noted that a similar analysis was performed also 
in Ref. \cite{KN-pS:IHJ}, 
in which a separable potential in momentum space was considered 
and $T$ matrix was obtained in a usual way.  
It is confirmed that our result has a similar tendency with their result. 
In particular, our result of NRv2 with $f_\pi=90$ MeV is found to be 
quantitatively the same as their result of Model B, as we summarize 
in Table \ref{PP-fp=90vsIHJ}. They found a lower pole as a virtual state 
in the energy independent case. However, we can not find it 
because the complex scaling method can not treat virtual states directly. 
In our energy-independent potential, there is a possibility that  
the lower pole may exist as a virtual state, not a resonance state. 

\begin{table}
\caption{Resonance of energy-independent and energy-dependent potentials. 
``E-dep.'' means the energy-dependent potential (NRv1 and 2, $f_\pi=110$ MeV) 
and ``E-indep.'' means its energy-independent version. 
 $Z_{H}$ and $Z_{L}$ are the complex energy of the higher and lower poles, respectively. ($Z=(M,-\Gamma/2)$) Both poles are obtained with the complex-range 
Gaussian basis ($\omega=2.0$) at $\theta=40^\circ$. 
All energies are in unit of MeV.
\label{PolePosition-fp=110}}
\begin{tabular*}{\textwidth}{l@{\hspace{0.1cm}}|@{\hspace{0.1cm}}cc|cc}
\hline
 &  NRv1  &       & NRv2   &  \\
 &  E-dep.&  E-indep. & E-dep. & E-indep.\\
\hline
$Z_H$ & (1416.6, $-19.5$)  & (1421.0, $-29.0$) & (1417.9, $-16.6$)  & (1418.5, $-23.8$) \\
$Z_L$ & (1380.6, $-133.4$) & ---               & (1395.2, $-137.9$) & --- \\
\hline
\end{tabular*}
\end{table}

\begin{table}
\caption{Resonance of energy-independent and energy-dependent potentials. 
Our result is obtained with NRv2 ($f_\pi=90$ MeV). Values of ``Model B'' 
are quoted from Table I in Ref. \cite{KN-pS:IHJ}. ``(V)'' in Model B indicates 
a virtual state. 
\label{PP-fp=90vsIHJ}}
\begin{tabular*}{\textwidth}{l@{\hspace{0.2cm}}|@{\hspace{0.2cm}}cc@{\hspace{0.6cm}}|@{\hspace{0.6cm}}cc}
\hline
 &  NRv2  &       & Model B \cite{KN-pS:IHJ}&  \\
 &  E-dep.&  E-indep. & E-dep.  & E-indep.\\
\hline
$Z_H$ & (1419.9, $-23.1$) & (1421.2, $-27.8$) & (1422, $-22$)  & (1423, $-29$) \\
$Z_L$ & ($\sim$1350, $\sim-66$) & ---               & (1349, $-54$) & (1325, 0) (V) \\
\hline
\end{tabular*}
\end{table}

\section{Summary and future plan}
\label{}

We have investigated a hyperon resonance $\Lambda(1405)$ 
which is well established to be a hadron-molecular state of 
$\bar{K}N$-$\pi\Sigma$ system with $s$-wave and $I=0$, 
using the coupled-channel Complex Scaling Method (ccCSM). 
Employing a {\it complex-range} Gaussian basis in the ccCSM and using 
a chiral SU(3)-based potential proposed in our previous work 
\cite{ccCSM-KN_NPA}, 
we have successfully confirmed the double-pole structure of $\Lambda(1405)$ 
as pointed out by many studies with chiral SU(3) models. 
In our previous study with real-range Gaussian basis \cite{ccCSM-KN_NPA}, 
we found only a single pole which corresponds to the higher pole of 
$\Lambda(1405)$.  
The complex-range Gaussian basis improves the description 
of the oscillatory behavior of continuum and resonance 
wave functions at large distance, since it involves trigonometric functions. 
Owing to this improvement, the eigenvalues of resonance poles with 
large decay width
 are clearly separated from the 2$\theta$ line indicating continuum states, 
although the separation of eigenstates is obscure 
for the real-range Gaussian basis. 
As a result, we have clearly obtained the lower pole as well as 
the higher pole. In case of a non-relativistic version 
of our chiral SU(3)-based potential, the lower pole is found 
at $(M, -\Gamma /2)=(1350, -66) \sim (1425, -163)$ MeV when 
the $f_\pi$ value runs from 90 MeV to 120 MeV. 
The eigenvalue of the lower pole depends strongly on the $f_\pi$ value, 
while that of the higher pole is quite independent 
to keep staying around (1420, $-20$) MeV. 

An advantage of the ccCSM is to obtain a wave function of a resonant state. 
We have confirmed that the ccCSM wave functions of both resonance poles 
exhibit the correct asymptotic behavior, comparing with 
the analytic form of resonance wave function in asymptotic region. 
Therefore, the ccCSM wave function obtained in our calculation 
is considered to be reliable numerically. 
Analyzing the ccCSM wave functions explicitly, 
we have revealed the nature of two resonance states just at the pole position 
on the complex-energy plane. 
The major component of the lower-pole state is found to be 
the $\pi\Sigma$ component, whereas the higher-pole state is dominated by 
the $\bar{K}N$ component, as former studies suggested that the lower 
(higher) pole couples strongly to the $\pi\Sigma$ ($\bar{K}N$) state. 
The $\pi\Sigma$ wave function of the lower pole  oscillates 
upto far distance, while the $\bar{K}N$ wave function of the higher pole 
is localized in a small distance. 
Thus, we conclude that the lower pole has qualitatively different nature from 
the higher pole. 

The ccCSM with the complex-range Gaussian basis is reliable on  
the identification of resonance poles and the analysis of their properties. 
It should be noted that we need a careful treatment for the very 
long-range behavior 
of the ccCSM wave function. In case of the lower pole, at very far distance 
(more than 50 fm) the ccCSM wave function shows the tiny amplitude, 
but deviates from the analytic form of 
the asymptotic wave function of the resonance. 
The tiny deviation of the order of  $10^{-5}$  
can affect the calculation of the meson-baryon mean distance numerically. 
Utilizing the asymptotic wave function together with the ccCSM wave function,  
we have obtained a stable value of the mean distance 
for the lower-pole state. The obtained mean distance is 
very small and has the imaginary part comparable to (in some case 
larger than) the real part; for example, $0.06-0.46i$ fm at $f_\pi=110$ MeV.
On the other hand, for the higher pole  
the mean distance has small imaginary value such as $1.4 - 0.3i$ fm. 
For a broad resonance such as the lower-pole state, 
it is difficult to interpret the meaning of the mean distance 
intuitively, even if its matrix element can be calculated. 

In order to know the origin of the double-pole nature, 
we have made further investigation with the ccCSM. 
As the strength of the $\bar{K}N$-$\pi\Sigma$ coupling potential is scaled, 
the energy position and property of both poles change continuously. 
When the coupling potential is completely switched off, 
the higher-pole state becomes a pure $\bar{K}N$ bound state, 
while the lower-pole state becomes a $\pi\Sigma$ resonance state. 
We have examined an energy-independent $\bar{K}N$ potential
which is constructed from our chiral potential by 
the static approximation. 
Then, only a single pole is found near the higher 
pole of the original energy-dependent potential, and 
no poles are found in the lower-pole region. 
Therefore, we conclude that the double-pole structure, 
especially the lower pole, is 
generated by the energy dependence of the chiral potential. 

Thus, the coupled-channel Complex Scaling Method has successfully 
revealed the double-pole nature of the $\bar{K}N$-$\pi\Sigma$ system 
which is involved by chiral SU(3)-based potentials. 
Through the present study, we expect that 
the ccCSM is quite a powerful tool to investigate resonant states 
of hadronic systems as well as nuclear system. 
Since the ccCSM can easily be applied to many-body system, 
we will tackle a three-body system ``$K^-pp$'' which is a prototype of kaonic nuclei. New experimental results on the $K^-pp$ are being reported from J-PARC \cite{Kpp-ex:JPARC-E15, Kpp-ex:JPARC-E27}. 
In a viewpoint of our study, the lower-pole state of $\Lambda(1405)$ 
as well as the higher-pole one is considered to couple to the $K^-pp$. 
It is interesting to investigate how the double pole affects  the $K^-pp$. 
The ccCSM is generally applicable to mesic nuclei other than kaonic nuclei. 
Since mesic nuclei have decay modes in usual case,  
most of mesic nuclei are considered to be a resonance state rather 
than a bound state, if they exist. We will investigate various 
kinds of mesic nuclei such as  $\eta$ \cite{eta:Nagahiro}, 
$\eta'$ \cite{etap:Nagahiro} and $\omega$ mesic nuclei.

\section*{Acknowledgements}

One of authors (A.D.) is thankful to Prof. T. Harada for 
his useful comments. This work was developed through several 
activities (meetings  and workshops) at the J-PARC Branch 
of the KEK Theory Center. 
This work is supported by JSPS KAKENHI Grant Number 25400286 
and partially by Grant Number 24105008.





\end{document}